\documentclass{article}

\usepackage{arxiv}

\usepackage[utf8]{inputenc} % allow utf-8 input
\usepackage[T1]{fontenc}    % use 8-bit T1 fonts
\usepackage{url}            % simple URL typesetting
\usepackage{booktabs}       % professional-quality tables
\usepackage{amsfonts}       % blackboard math symbols
\usepackage{nicefrac}       % compact symbols for 1/2, etc.
\usepackage{microtype}      % microtypography
\usepackage{graphicx}

\usepackage{hyperref}
\usepackage{doi}

\title{Security and Privacy for Low Power IoT Devices on 5G and Beyond Networks: Challenges and Future Directions}

\author{Jonathan Cook*,
        Sabih ur Rehman*
        and M. Arif Khan*\\ % <-this % stops a space\\
        {*School of Computing, Mathematics and Engineering, Charles Sturt University, Australia}}
\begin{document}
\maketitle

\begin{abstract}
	The growth in the use of small sensor devices, commonly known as the Internet of Things (IoT), has resulted in unprecedented amounts of data being generated and captured. With the rapidly growing popularity of personal IoT devices, the collection of personal data through such devices has also increased exponentially. To accommodate the anticipated growth in connected devices, researchers are now investigating futuristic network technologies that are capable of processing large volumes of information at much faster speeds. However, the introduction of innovative network technologies coupled with existing vulnerabilities of personal IoT devices and insufficient device security standards is resulting in new challenges for the security of data collected on these devices. 
While existing research has focused on the technical aspects of security vulnerabilities and solutions in either network or IoT technologies separately, this paper thoroughly investigates common aspects impacting IoT security on existing and futuristic networks, including human-centric issues and the mechanisms that can lead to loss of confidentiality.
By undertaking a comprehensive literature review of existing research, this article has identified five key areas that impact IoT security for futuristic next generation networks. Furthermore, by extensively analysing each area, the article  reports on conclusive findings and future research opportunities for IoT privacy and security for the next generation of network technologies.
\end{abstract}

% keywords can be removed

\section{Introduction}\label{Introduction}
Personal security is being challenged in new ways since the introduction of personal IoT devices. In the last decade, 
 new developments and more affordable devices has resulted in the number of personal IoT devices increasing significantly. 
 While innovative IoT devices have grown in number and network performance has increased, personal IoT devices have also  introduced significant privacy and security challenges \cite{arias_wurm_hoang_jin_2015, bhushan_agrawal_2019, thierer_2014}. With an ever-growing list of personal devices entering circulation, users of personal IoT devices face a loss of confidential information with significant security implications from a range of issues \cite{thierer_2014}, many of which have uncertain outcomes on emerging network technologies beyond the fifth-generation (5G) wireless spectrum. Through a systematic review, this paper will identify risks and causes of the loss of confidentiality and security from personal IoT devices. It will additionally address these concerns from an Australian context, investigating personal IoT device security standards.

 \begin{figure*}[t]
		\centering
		\includegraphics[width=0.9\linewidth]{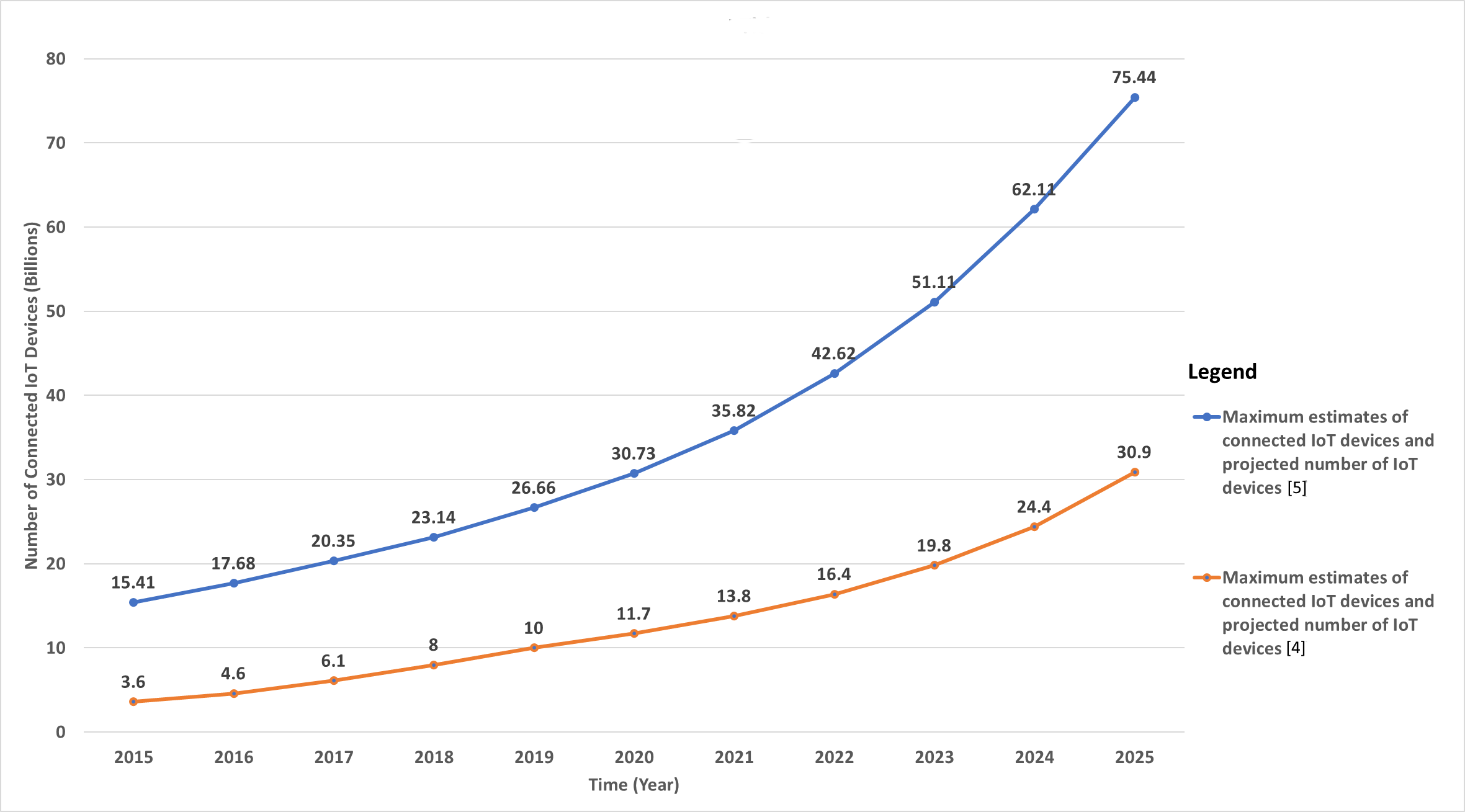}
		\caption{Number of IoT device connections in billions by 2025 with upper and lower bound estimates - \cite{djenna_harous_saidouni_2021} and \cite{alam_2018}}
		\label{fig:estimated_iot_devices}
	\end{figure*}

 As highlighted in the previous paragraph, IoT has exploded in popularity in recent years. This growth has been driven by new consumer products that autonomously collect and transmit data. The growth in access to increasingly affordable personal IoT devices has allowed consumers to gain insights into their health and fitness, improve efficiency, and automate tasks, which results in an overall better quality of life. Although IoT has grown in popularity in recent years, it is important to develop an understanding of the technology and how it will impact futuristic networks. This interpretation will allow for a better understanding of the challenges to privacy and security that this technology introduces.  
 IoT, as a technology, is defined as a global network of connected sensors and actuators that communicate autonomously, that is, without human intervention \cite{atzori_iera_morabito_2010}. The term was originally conceived by Kevin Ashton of Procter \& Gamble when he first envisioned IoT using communication between Radio Frequency Identification (RFID) devices and the internet in the company's supply chain in 1999 \cite{rayes_salam_2016}. Despite the innovative concept, the widespread application of IoT in commercial, government and private usage did not eventuate until a decade later.

 Since 2010, IoT has experienced exponential growth in worldwide usage, with estimates of the number of connected devices anticipated to be between 30 billion \cite{djenna_harous_saidouni_2021} and 75 billion \cite{alam_2018} by 2025, as shown in Figure \ref{fig:estimated_iot_devices}, with the upper and lower estimates of connected devices illustrated. Consequently, the growth in the use of IoT has resulted in unprecedented amounts of data being generated and captured. With the emergence of 5G and beyond networks (5GBN), such as sixth-generation (6G) technology, research has already commenced developing new technological architectures that can process the large volumes of information that is captured via these devices at much faster speeds. 
 These advances, however, come with significant risks to the personal security and privacy of all users of personal IoT devices connected via such networks. This paper will take a deep dive into existing literature to investigate the threats to privacy and security originating from such devices and the development of protective mechanisms to enhance personal privacy and security.

	\subsection{Problem Domain}
\begin{figure*}[t]
		\centering
		\includegraphics[width=0.8\linewidth]{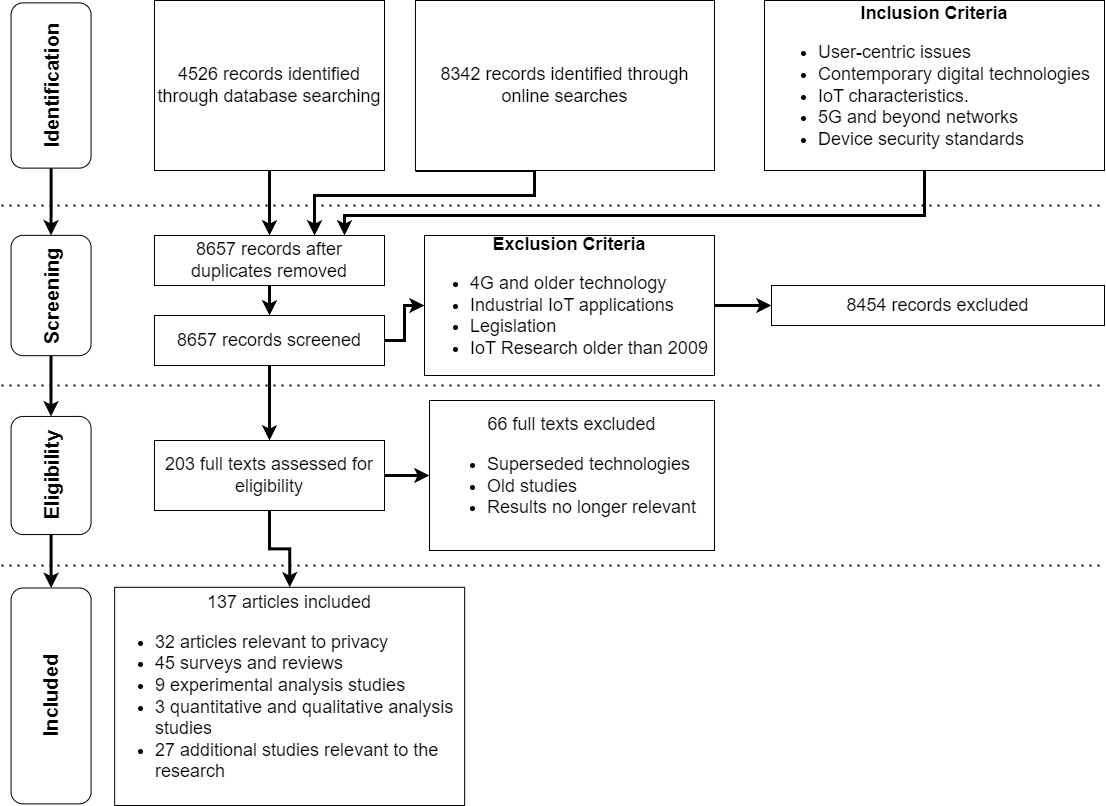}
		\caption{Literature selection process}
		\label{fig:litrev_flow}
	\end{figure*}

	An examination of the existing literature has demonstrated that while additional technologies can be utilised to enhance IoT security, the proposed methods do not fully address the three principles of security being confidentiality, integrity and availability \cite{rayes_salam_2016}. Additionally, an absence of device security standards across the industry contributes to an increased risk of loss of privacy and security. Consequently, privacy concerns, which are paramount in cyber security protection, have been absent in 5GBN IoT research, particularly in the Australian context. From this literature examination, an exhaustive list of relevant high-quality journal articles
 have been analysed to investigate the current field of knowledge and concerns researchers have identified. Due to the rapidly evolving landscape of IoT and cybersecurity, this paper investigated additional research areas, including industry white papers and government publications for the most recent developments. The process of article selection is highlighted in Figure \ref{fig:litrev_flow}.

 An extensive analysis of existing literature has revealed that the primary method to enhance the privacy and security of personal IoT devices on 5GBN has been to utilise cutting-edge contemporary digital technologies such as blockchain, Machine Learning (ML), and Artificial Intelligence (AI). Many of the scrutinised articles focus on adding protective layers, such as ML, without fully addressing all IoT characteristics and the three basic principles of security which are \textit{confidentiality}, \textit{integrity} and \textit{availability} concurrently \cite{pfleeger_sharilawrencepfleeger_margulies_2015}.
 While recent studies, such as those conducted by the authors in \cite{salahdine2023security} and \mbox{\cite{kumar2022security}} explore the use of emerging technologies and threats, human-centric issues and the mechanisms that lead to data exploitation through legitimate means are not investigated.
 Consequently, the preliminary investigation has identified the following issues related to cyber security vulnerabilities surrounding IoT data privacy concerns.

	From the preliminary investigation, this research identified several issues affecting personal IoT privacy and security. 
\ Individual users of personal IoT devices are an unreliable and uncontrollable variable of data confidentiality.
 A reliance on end-users of IoT devices to apply strong password policies and ensure that device software is kept up to date is inadequate and can lead to significant security concerns \cite{vaniea_rashidi_2016, tawalbeh_muheidat_tawalbeh_quwaider_2020}.  Although the safe handling of confidential information is paramount to ensuring the security of IoT devices on 5GBN \cite{varadharajan_bansal_2016}, in the study by the authors of \cite{subahi_theodorakopoulos_2018}, half of the device manufacturers investigated did not adopt an adequate privacy policy for their devices. Further, two in the study failed to comply with their own stated policies. Data security concerns are heightened as data may not always be secure. Failures of data security are, in part, the result of the low power consumption and limited processing capacity of IoT devices. These power constraints often result in abandoned or weaker encryption methods \cite{khatib_ahmed_kamaleldin_abdelghany_mostafa_2018}. Due to these issues, IoT on 5GBN will increasingly rely on emerging technologies, such as AI and ML, to enhance security and privacy, which are prone to cyber security threats \cite{ghorbani_mohammadzadeh_ahmadzadegan_2020}. 
 While researchers continue to investigate the use of these innovative technologies to enhance security and privacy preservation, the resource-intensive nature results in the technologies being deployed at the edge, either in the cloud or fog.
 However, the absence of physical layer security and IoT device security standards are problematic and create challenges for manufacturers to ensure their devices are secure by design. As a result of the investigations, this paper has identified significant issues inhibiting the development of IoT security on futuristic networks, as summarised below:
 
 \begin{itemize}
	\item Studies have demonstrated that individual users often fail to follow recommended security procedures for data protection.

	\item Half of IoT device manufacturers investigated in one study did not adopt an adequate privacy policy for their devices.

	\item IoT device characteristics contribute to data security concerns due to low power consumption and limited processing capacity.
	
	\item The physical layer of IoT remains vulnerable to adversarial attack due to the limited computational processing power.

    \item To enhance the security of data collected by personal IoT devices, the use of contemporary digital technologies is often called upon, which are vulnerable to cyber security threats.

\end{itemize}

\begin{figure*}[t]
		\centering
		\includegraphics[width=0.96\linewidth]{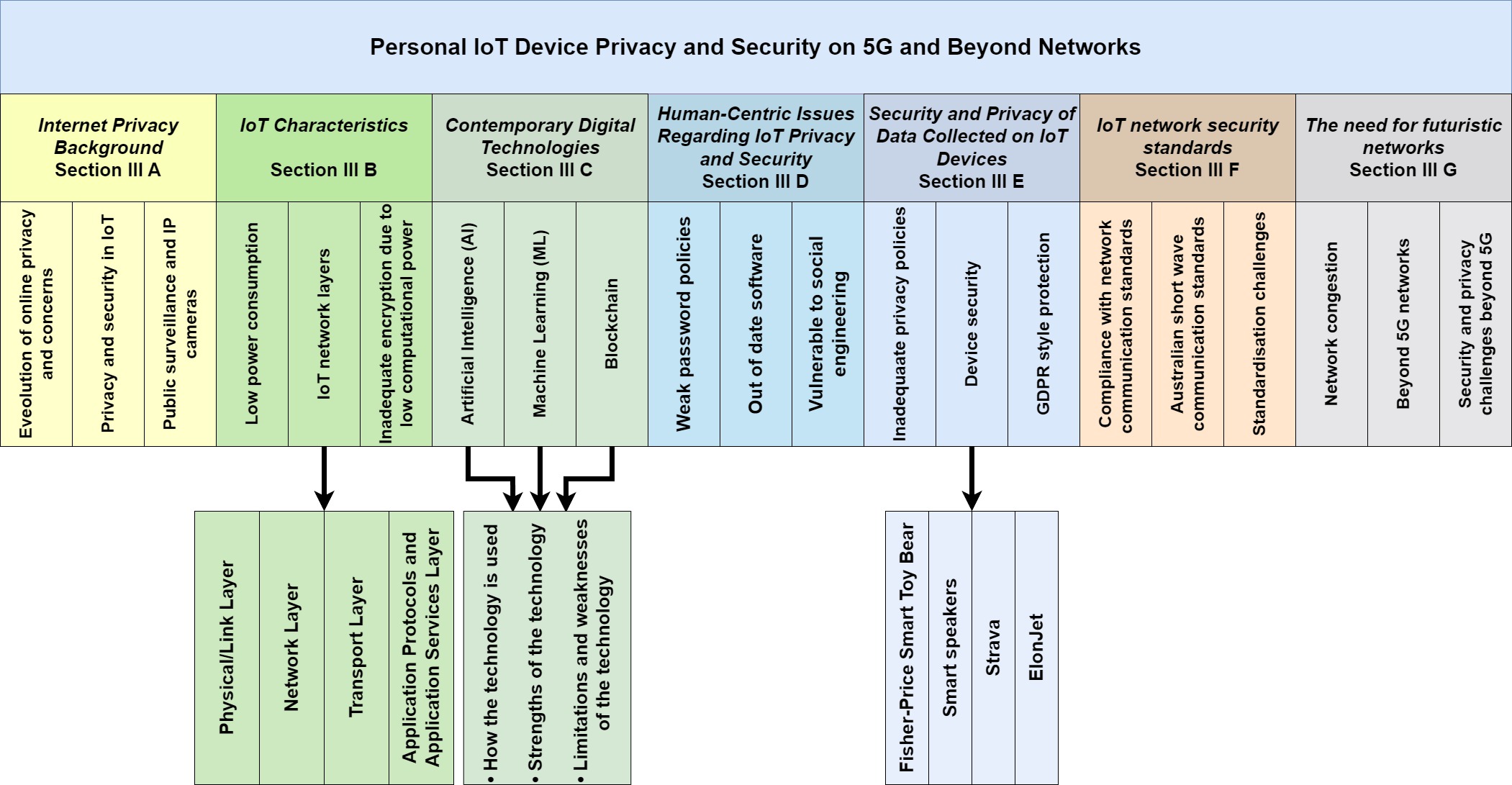}
		\caption{Categorical organisation of this review}
		\label{fig:article_flow}
	\end{figure*}

\subsection{Main Contributions of this Paper}

With the key challenges identified above, this paper has identified future research directions to enhance personal IoT device security.
 The main contributions of this article are summarised as follows:

\begin{itemize}
    \item Identify five key areas that impact IoT security for futuristic next generation networks through a systematic literature review.
    \item Classify the risks to security and privacy for users of personal IoT devices on next generation wireless networks.
    \item Highlight security risks associated with an absence of IoT device security standards on futuristic networks.

\end{itemize}

\subsection{Structure of the paper}
The remainder of this manuscript is organised as follows: Section \ref{Methodology} explains the methodology used and defines the research questions and article selection process. 
Section III will categorically investigate each of the research areas that influence privacy and security on 5GBN, with the organisation of the review illustrated in Figure \ref{fig:article_flow}.
Section \ref{HistoryIOTPrivacy} explores the evolution of internet privacy and how device exploitation has infiltrated personal privacy and security.
Section \ref{IoTCharacteristics} investigates how IoT characteristics contribute to security vulnerabilities.
%Section \ref{IoTCharacteristics} investigates the role of IoT characteristics in the security and privacy of the data collected by personal IoT devices and the risks to the data collected by such devices.
Section \ref{ContemporaryDigitalTechnologies} examines the role of contemporary digital technologies in securing personal security and the risks of over-reliance on these technologies. 
Section \ref{HumanCentricIssuesRegardingIoTPrivacySecurity} explores how human-centric issues generate security vulnerabilities for personal IoT devices and the challenges these issues create in security IoT on futuristic networks.

Section \ref{SecurityPrivacyData} investigates how data is collected and managed by personal IoT devices and the risks associated with the mismanagement and misuse of the data. Section \ref{IotSecurityStandards} explores the role of IoT security standards in enhancing personal IoT device security. In Section \ref{Beyond5G}, this paper investigates the need for futuristic network technologies and the potential implications for personal IoT device security. Section \ref{FurtherResearch} presents the key findings and future research opportunities of this literature review. Finally, Section \ref{Conclusion} presents the conclusion of this work. 

\section{Methodology} \label{Methodology}

This literature review identifies the challenges, solutions, and future work to enhance privacy and security for personal IoT devices on 5GBN. To undertake this review, a preliminary investigation was undertaken to identify gaps in the existing knowledge of personal IoT security on 5GBN. From the preliminary investigation, issues regarding privacy, physical layer security and an absence of security standards were identified as areas requiring further investigation.

\subsection{Inclusion and exclusion criteria} \label{InclusionExclusionCriteria}

	With an area of research identified, the development of inclusion and exclusion criteria for the article analysis was developed. The development of inclusion and exclusion criteria for the research allows it to remain focused and on topic. It additionally aids with the development of database search terms to identify relevant literature. The exclusion criteria are listed below with a brief explanation for the exclusion:
	\begin{itemize}
		\item Limit the years of research from 2009 until 2022. As the IoT industry is rapidly evolving \cite{ tiwari_sharma_kaushik_tiwari_bhushan_2019}, it is necessary for this research to limit the years of research for specific areas to no older than 2009. As highlighted by the authors of \cite{son2019collaborative}, they make a case that the age of IoT began between 2008 and 2009. Although the term was created much earlier, it was during this time that the number of devices surpassed the number of people. Additionally, the network technology in 2009 was Third Generation (3G) wireless technology, with 5G only entering service in 2019. With 3G being superseded and approaching the end of life, it was necessary to define a point where IoT in personal communication was beginning to enter mainstream society and the network technology available at the time is still accessible by modern devices.
 
		\item 4G and older technology. As the research is focused on 5GBN technology, the inclusion of redundant networking technology would add little value to the research.
		\item Industrial IoT applications. Although related, personal and industrial IoT device security should be investigated separately.
		\item Legislation. While security and privacy encompass areas of law, it is beyond the scope of this research to identify legal requirements for the use of IoT and network communication across multiple jurisdictions.
	\end{itemize}

 The list of inclusion criteria is included below and allows for the deep analysis of the research questions formulated in Section \ref{ResearchQuestions}:
	\begin{itemize}
		\item User-centric issues. As users of low-power IoT Devices have been identified as a significant security challenge for IoT data, it is necessary to investigate the role of individual users.
		\item Contemporary digital technologies. With a growing reliance on contemporary technologies such as blockchain, AI and ML, the inclusion of these technologies is needed to investigate how they can enhance protection and identify potential issues from their use.
		\item Security and privacy of data collected on IoT devices. As this research focuses on data privacy and security of IoT devices on 5GBN, it is necessary to include it in the research,
		\item IoT characteristics. As IoT characteristics are contributing factors to their overall security, the inclusion of IoT characteristics is necessary. 
		\item 5G and beyond networks. While researchers are currently investigating 5G applications, preliminary investigations of futuristic networks are already underway. The inclusion of relevant next-generation technologies will assist in identifying challenges that lie ahead.
        \item Personal IoT device security standards. The purpose of standards is to ensure the safety of products, services, and systems through setting specifications, procedures, and guidelines \cite{standardsaustralia_2010}. An investigation of security standards for personal IoT devices will help to identify areas that need further development.
	\end{itemize}

 \subsection{Research Questions} \label{ResearchQuestions}

 The following initial thought-provoking research questions have been investigated with the research issues identified above. In addressing these questions, the research has developed an understanding of \textit{personal IoT device security and the role users play in  personal IoT device security on futuristic networks}. Additionally, the research has explored what data may be collected, how it is used and how \textit{personal IoT device security is enhanced using contemporary digital technologies}. Finally, the research explored gaps in existing knowledge that impact IoT security and privacy on futuristic networks. In summary, the following research questions (RQs) have been explored as a part of this study:
 
 (RQ1): \textit{How is privacy protected on low-power personal IoT devices? } \\
 (RQ2): \textit{What is the impact of IoT characteristics on privacy for personal IoT devices?} \\
(RQ3): \textit{What is the role of contemporary digital technologies in privacy preservation on futuristic networks?}	\\
 (RQ4): \textit{How do existing standards protect confidential data originating from low-powered personal IoT devices?} \\
	 
	\subsection{Search Process} \label{SearchProcess}
	To answer the research questions outlined above, a comprehensive search process was conducted to capture the relevant literature. In aligning with the inclusion and exclusion criteria outlined in Section \ref{InclusionExclusionCriteria}, literature searches were kept within the period of 2009 until 2022. However, as part of the background discussion of internet privacy, which forms an integral part of IoT privacy, the research included journal articles from 1989. Google Scholar was used to identify thematic trends within articles within the periods identified. From the papers identified, the titles, keywords and abstract were collected for thematic analysis. From this search process, we were able to identify 137 articles for inclusion, of which the most related articles are shown in Table \ref{tab:related-literature}. The search terms used in Google Scholar are highlighted as follows:

  TITLE\_ABS\_KEY((“IoT privacy” OR “IoT security” OR “network privacy” OR “network security” OR “internet privacy” OR “internet security” OR “IoT AI/ML” OR “IoT machine learning” OR ”IoT artificial intelligence” OR “IoT blockchain” OR “IoT security standards”) AND (“5G” OR “6G” OR “futuristic networks” OR “beyond 5G” )) AND (LIMIT\_TO (PUBYEAR, 2022) OR LIMIT\_TO (PUBYEAR, 2021) OR LIMIT\_TO (PUBYEAR, 2020) OR LIMIT\_TO (PUBYEAR, 2019) OR LIMIT\_TO (PUBYEAR, 2018) OR  OR LIMIT\_TO (PUBYEAR, 2017) OR LIMIT\_TO (PUBYEAR, 2016) OR LIMIT\_TO (PUBYEAR, 2015) OR LIMIT\_TO (PUBYEAR, 2014) OR LIMIT\_TO (PUBYEAR, 2013) OR LIMIT\_TO (PUBYEAR, 2012) OR LIMIT\_TO (PUBYEAR, 2011) OR LIMIT\_TO (PUBYEAR, 2010) OR LIMIT\_TO (PUBYEAR, 2009)) AND (LIMIT\_TO (DOCTYPE, “cp”) OR LIMIT\_TO DOCTYPE, “ar”)  OR LIMIT\_TO DOCTYPE, “ch”)   OR LIMIT\_TO DOCTYPE, “bk”) ) AND (LIMIT\_TO) (LANGAUGE, “English”))

  %%
% table* to table begin and end
%%

\begin{table*}
  \caption{Related literature}
  \tiny
  \label{tab:related-literature}
  \begin{tabular}{|p{2cm}|p{3cm}|p{3cm}|p{3cm}|p{3cm}|}
  \hline
  \textbf{Publications} &
    \textbf{Methodology} &
    \textbf{Strength} &
    \textbf{Limitations} &
    \textbf{Future Work} \\ \hline
  \cite{sicari_rizzardi_coen-porisini_2020} &
    Survey of existing measures to   enhance privacy and security &
    In depth analysis of current   security and privacy protocols &
    Focus on existing network technology and the current IoT landscape &
    Energy conservation security   measures, low latency guarantee, low overheads \\ \hline
  \cite{tawalbeh_muheidat_tawalbeh_quwaider_2020} &
    Survey of existing risks and   development of new layer framework to enhance security &
    In depth analysis of existing   risks and framework development for additional study &
    Focus on previous and existing network   technology &
    Cryotographic security methods   that can work efficiently on IoT devices and standardised data collection   method \\ \hline
  \cite{lin_srivastava_zhang_djenouri_aloqaily_2021} &
    Experimental development of a   probabilistic technique to enhance security &
    Identifies that privacy and   security will be vulnerable targets, especially to Man-in-the-middle attacks,   for emerging 6th generation IoT technology. &
    Reliance on third-party sources   for security and inefficiency &
    Focus on edge computing to remain   efficient \\ \hline
  \cite{nguyen_ding_pathirana_seneviratne_li_niyato_dobre_poor_2021} &
    Survey of benefits of 6G   technology for IoT &
    In depth analysis of current and   future benefits of 6G technology &
    Limited research on privacy and   security &
    Investigation of energy   efficiency issues relating to 6g IoT networks \\ \hline
  \cite{lee_2020} &
    Empirical analysis of 265   samples  measuring differences in   vulnerability factors, along with privacy concerns &
    In depth analysis on IoT privacy   highlighting user vulnerability as the highest impact on home IoT privacy   concerns &
    No research conducted with   reference to 5GBN technology &
    Research analyses of external   threats \\ \hline
  \cite{menard_bott_2020} &
    Quantitative based analysis of   user and IoT device usage with privacy modelling &
    Investigation into understanding   of consumers privacy concerns with IoT &
    Research on 5GBN technology   not undertaken and limited to android users &
    Similar work of iOS users and the   inclusion of 5GBN technology implementations \\ \hline
  \cite{psychoula_singh_chen_chen_holzinger_ning_2018} &
    Qualitative based research using   online surveys and interviews &
    Analysis of users perceptions and   willingness to forego privacy in favour of IoT services &
    Limited to a small number of   participants and no focus on emerging networks &
    Which factors of trust, risk,   perception, knowledge, awareness or all determine privacy attitude among   users \\ \hline
  \cite{seliem_elgazzar_khalil_2018} &
    Survey of existing measures for   privacy preservation in IoT environments &
    In depth analysis of current   security and privacy protocols and limitations &
    No research conducted with   reference to futuristic network technology &
    Application to emerging network   technologies \\ \hline
  \cite{liranzo_hayajneh_2017} &
    Experimental analysis of the   privacy and security risk of IoT IP Cameras &
    Demonstrates risks to privacy   associated with IP cameras &
    Limited to IP cameras and ignores   emerging networks &
    Investigate how to protect   consumer privacy and enhance IP camera security \\ \hline
  \cite{imoize_adedeji_tandiya_shetty_2021} &
    Comprehensive survey of existing   threats and proposed measures to enhance privacy and security &
    Focused on 6G technology and   examines multiple different proposals &
    Does not address privacy for the   public &
    Connectivity in the 6G era and   beyond, including channel estimation, security, and underwater communication. \\ \hline
  \cite{moncrieff_venkatesh_west_2009} &
    Survey of privacy concerns   regarding public surveillance &
    In depth analysis of privacy   implications for the public through surveillance &
    Pre-dates IoT and modern network   security concerns &
    Review of privacy concerns from   IP cameras and modern IoT devices on modern networks \\ \hline
  \cite{sivaraman_gharakheili_fernandes_clark_karliychuk_2018} &
    Experimental testing and   qualitative analysis of IoT devices &
    In depth analysis of common IoT   monitoring devices and the privacy implications they pose &
    No research conducted with   reference to futuristic network technology &
    Renewed testing on modern devices   and networks \\ \hline
  \cite{wang_zhu_zhang_zhang_yu_zhou_2020} &
    Review of security and privacy In   6G networks &
    In depth analysis of security and   privacy concerns in 6G networks which focuses on connected technologies to   counter threats &
    Limited discussion on   implications for IoT on futuristic networks beyond 5G technology &
    Additional research on AI   integration in relation to cybersecurity threats identified by other   research \\ \hline
  \cite{siriwardhana_porambage_liyanage_ylianttila_2021} &
    Deep analysis of the use of artificial   intelligence in 6G networks &
    Identifies opportunities and   challenges of 6G networks and proposes the use of artificial intelligence as   a tool to enhance security and privacy &
    Very little investigation   regarding IoT   security concerns &
    While the author discussed   ethical issues, additional research is required to understand the   implications to privacy \\ \hline
  \cite{nguyen_lin_cheng_hwang_lin_2021} &
    Review of privacy and security in   6G networks &
    In depth analysis of technologies   and challenges to security on 6G networks &
    Absence of IoT based technologies   and security challenges in the research &
    Additional research in   authentication protocols in relation to IoT to enhance privacy on 6G networks \\ \hline
  \cite{mrabet_belguith_alhomoud_jemai_2020} &
    A survey of IoT security threats   based on IoT architecture &
    In depth analysis of security   threats on each layer of IoT architecture &
    No discussion on the development of futuristic network technology and the implication of developing networks on IoT security &
    Application of the proposed   classification on 6G networks \\ \hline
  \cite{harkin_mann_warren_2022} &
    A qualitative based survey of IoT regulations in Australia &
    An extensive study on the regulation of IoT in Australia &
    No discussion on futuristic networks and a focus on regulation &
    Calls for further IoT regulation in Australia due to no clear regulation direction \\ \hline  
  \cite{internetofthingpaper_2020} &
    A survey on the existing standards of IoT in Australia &
    A government regulators survey of IoT security and standards applications in Australia which identifies weaknesses and directions for improvement &
    Limited discussion on futuristic networks and fails to recommend a solution to compliance challenges. &
    Continued monitoring of standards development, monitoring spectrum demand and updating licensing arrangements to support IoT as required \\ \hline 
  \cite{salahdine2023security} &
    A survey of security and privacy   challenges on 5G and beyond networks &
    Identifies several factors contributing to weaker security on 5G and beyond networks, including an absence of security standards &
    Very limited discussion on IoT and does not address human-centric issues regarding security. Discussion on regulation implementation &
    Encourages enhancing IoT security to help secure 5G networks and future investigation into a proposed OFDM-SIS scheme for secrecy performance \\ \hline  
  \cite{kumar2022security} &
    A review of 5G and IoT security   challenges &
    Discusses different settings of IoT and the security challenges associated with their application. Identifies challenges with the architecture and technologies &
    Does not investigate security standards or regulations and no research on human-centric issues related to security &
    Identifies a range of challenges primarily associated with contemporary digital technologies that require future investigation \\ \hline  
  \end{tabular}
\end{table*}

  \section{IoT in Personal Communication} \label{IoT in Personal Communication}

   This section explores the multiple facets that comprise the security and privacy of IoT in personal communication and the implications for personal IoT security with the anticipated arrival of futuristic network technology. Upon completing this section, this paper will have identified the leading causes of security vulnerabilities for personal IoT devices, emerging technologies to enable strengthened security, the role of security standards in protecting users of personal IoT devices and potential hazards with the arrival of futuristic network technologies.

\subsection{Evolution of Internet Privacy} \label{HistoryIOTPrivacy}
	One of the earliest endeavours of privacy research on computer networks was conducted in 1989 \cite{bishop_1991}. At the time, the privacy concerns specifically regarded the most common form of network communication being electronic mail, otherwise known as email. The author of the paper identified a lack of security mechanisms for online communication and proposed a range of measures, such as encryption, to enhance security and protect privacy. Privacy concerns originally identified in 1989 continued as an area of investigation for a decade, emerging as the biggest concern facing users of the internet in 1999 \cite{benassi_1999}. At the time, privacy concerns among internet users surpassed other issues such as spam, ease of use, cost and even security. 

 Although the concept of IoT was first introduced in 1999, around the same time as the privacy research by the authors of \cite{benassi_1999}, IoT technology pre-dates many of the technologies discussed in the authors study.
 With the widespread use of the internet in its infancy, the new concept of IoT, and initial research of 4G network technology which would become commercially available a decade later, the privacy and security concerns of these converging technologies remained uncertain \cite{abioye_joseph_ferreira_2015}.
	
	The introduction of widely used social networks and popularity of services provided by search engines such as Google in the early part of the twenty-first century heralded a new era of privacy issues for online citizens. Interactions online and between website users, their social connections online, their location and activities, were soon avenues of exploitation by both website administrators and cybercriminals. Public awareness to privacy issues arising through unregulated social networks, search and general website browsing were a driving force for the introduction of measures to allow users to protect some of their details online. However, much of the data collected in the early 2000’s originated through direct interaction with websites or through tracking of activities using Cookies. The ability of IoT to collect data without direct interaction altered the dynamics of online privacy and security. 
	
	Since the research by the authors in 1999 \cite{benassi_1999}, the widespread usage of the internet has grown exponentially \cite{stanton_2019}, becoming a fundamental part of society, driving global e-commerce \cite{babenko_kulczyk_perevosova_syniavska_davydova_2019}, industry and social interactions \cite{arora_bansal_kandpal_aswani_dwivedi_2019}. With an estimated figure of approximately 5.5 billion users at the end of 2022 \cite{internetworldstats_2019}, its usage encapsulates a significant proportion of global citizens. The growth in its use has partly been driven by increased access to affordable, portable devices with internet connectivity, such as smart mobile phones and wearable devices \cite{rehman_liew_wah_shuja_daghighi_2015}. However, while increased access to internet connected devices has allowed more people to engage online, the growth of IoT devices has resulted in most of the internet traffic having no direct human interaction \cite{labovitz_iekel-johnson_mcpherson_oberheide_jahanian_2010}. While on the surface a lack of human initiated internet engagement would appear to have a negligible impact on privacy, the increased amounts of data collection by IoT devices that have the ability to collect sensitive personal information are proving problematic for privacy preservation  \cite{ziegeldorf_morchon_wehrle_2013}. With ever-increasing numbers of personal IoT devices, privacy preservation is becoming increasingly challenging for the average person \cite{mehmood_natgunanathan_xiang_hua_guo_2016, caron_bosua_maynard_ahmad_2016}, and for cyber security researchers tasked with enhancing IoT security.

 While network communication, privacy and security are now considerably more sophisticated than they were in 1989, the development of IoT in conjunction with emerging futuristic wireless network technology is creating not only new challenges for researchers, but also new avenues for exploitation. While online security and privacy has been an area of constant investigation, the potential for device exploitation, particularly through public surveillance which exposes users’ privacy, has been absent from research. Additionally, the vast amount of data collected by such devices is creating new security challenges for cyber security researchers.
 
 In summary, privacy preservation through public surveillance was initially investigated by the authors of \cite{moncrieff_venkatesh_west_2009} in 2009. In this study, the authors primarily investigated the impacts of surveillance from cameras and the impacts they have on privacy. Their study, while addressing many concerns regarding privacy preservation from public surveillance, failed to consider the future potential of IoT devices and the risks they pose to privacy, only briefly addressing IP connected cameras. While IoT was new at the time and 3G network technology was the mainstay of wireless communication \cite{alsharif_kelechi_albreem_chaudhry_zia_kim_2020}, failing to fully address potential privacy and security concerns originating from mass data collection could be considered an oversight. Additionally, privacy exploitation through technologies available and widely used at the time was also not vigorously investigated. One such example is the Global Positioning System (GPS) location tracking through smart devices such as mobile phones. The concept of location tracking through GPS and wireless technology was investigated in 2002 \cite{bajaj_ranaweera_agrawal_2002}, demonstrating how GPS data can be transmitted across wireless networks. Nevertheless, while the research conducted by the authors of \cite{moncrieff_venkatesh_west_2009} does not consider IoT's implications on privacy, their investigation lays the foundations for privacy preservation in computing networks.

\subsection{IoT Characteristics} \label{IoTCharacteristics}
	The characteristics of IoT form a leading role in the cyber security and privacy challenges of personal IoT devices on 5GBN. These challenges stem from a requirement for low power consumption, which subsequently results in low computational processing power. Following these limitations at the device level, the network communication characteristics of IoT create additional challenges for IoT security. This paper will now discuss each of the characteristics in detail.

  %  \begin{figure*}[t]
	%	\centering
	%	\includegraphics[width=0.9\linewidth]{figures/iot_stack.jpg}
%		\caption{IoT architecture - \cite{rayes_salam_2016}}
%		\label{fig:iot_layers}
%	\end{figure*}

 \subsubsection{Low Power Consumption and Limited Processing Resources}

 One of the prevailing requirements for IoT devices is the need for low power usage \cite{hussain2023efficient}. This is often due to the need for devices to operate by battery power and are required to run for a significant time between charging cycles \cite{henkel_pagani_amrouch_bauer_samie_2017}. The low power consumption of personal IoT devices forms security and privacy challenges that can inhibit privacy preservation and security from the device \cite{hameed_khan_hameed_2019}.
	As personal IoT devices operate for prolonged periods on battery power, energy conservation is a priority to ensure the device remains operational between charging cycles. 
	This energy conservation creates a trade-off. To conserve power, IoT computational processing power is commonly restricted, exposing a flaw in the security of IoT devices \cite{rayes_salam_2016}. 

 \subsubsection{Processing Power Limitations}\label{ProcessingPowerLimitations}

 As briefly mentioned above, security is often sacrificed for the benefits of low power usage, potentially exposing sensitive information and weaknesses in the device's network \cite{rayes_salam_2016}. The authors of \cite{khatib_ahmed_kamaleldin_abdelghany_mostafa_2018} identified this limitation and proposed a solution for encryption while limiting the power consumption of low powered devices. While the solutions of encryption suit the requirements of an almost instantaneous response time, the authors of \cite{khatib_ahmed_kamaleldin_abdelghany_mostafa_2018} recognise that their proposal may result in less secure encryption techniques as device power management becomes active. 
	
	The paper by the authors of \cite{tsai_leu_you_chang_hu_park_2019} proposed a different solution for encryption that uses AES 256 encryption, promising to enhance cryptographic security without significantly compromising power management features. The authors propose a Low-Power AES Data Encryption Architecture (LPADA) that maintains the cryptographic security of AES encryption but maintains the low power usage requirements by encrypting and decrypting data using low-power SBox, power gating and power management techniques. While the model proposed by the authors of this study results in significant reductions in power, the study authors recognise additional research needs to be conducted to further reduce encryption power consumption. However, with battery life and efficiency improving \cite{mandyam_2010} and devices increasing in processing power due to more powerful microprocessors \cite{ojo_giordano_procissi_seitanidis_2018}, the argument for embedded cryptography is long overdue. 

    Recently, a solution to add cryptography at the IoT device level was  proposed by the authors of \cite{ahmed2022cryptographic}. In their investigation, the authors devised a solution for end-to-end encryption in IoT healthcare on 5GBN. The solution proposed by the authors uses a two-layer symmetric encryption for the data before transmitting it via software-defined network (SDN) routers. While the proposed solution provides encryption at the device level, the authors note that the strength of encryption can be enhanced by adding more key layers. However, as they identify, more key layers increases the cost of encryption and consumes more of the limited available memory of the IoT device. A proposed solution by the authors is to apply additional encryption key levels based on the data's sensitivity. However, as we demonstrate in Section \ref{SecurityPrivacyData}, even seemingly harmless personal data can be mined and exploited using such an approach.

  While cryptography is one area under investigation to enhance the physical layer security of IoT, researchers are exploring other solutions to solve the security concerns of the physical layer. The authors in \cite{sharma2023increasing} propose using Visible Light Communication (VLC) through hyperchaos-based security measures as one possible security enhancement. Their research notes that VLC has higher security than traditional radio frequency communication. When coupled with hyperchaos, it significantly enhances security, particularly concerning eavesdropping. However, a limitation of VCL hyperchaos is the direct line of sight required for communication and the inability to penetrate physical objects such as walls. This limitation reduces the effective practical use of VLC in a wide range of applications where line-of-sight communication is not guaranteed.

  \begin{figure*}[t]
		\centering
		\includegraphics[width=0.9\linewidth]{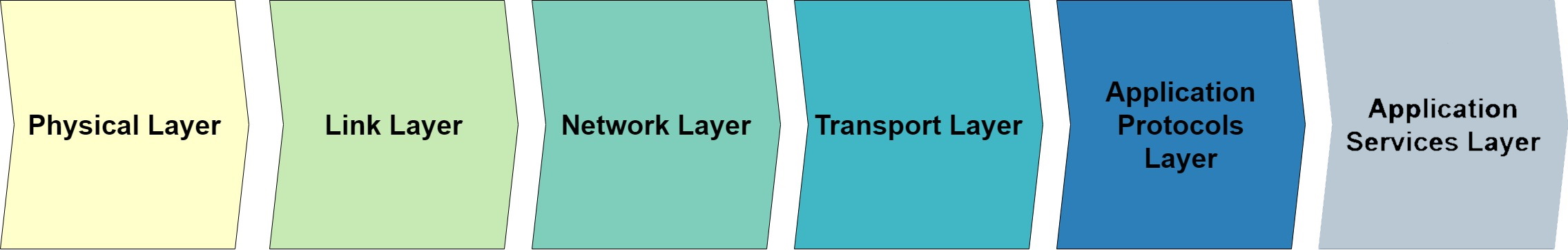}
		\caption{IoT architecture - \cite{rayes_salam_2016}}
		\label{fig:iot_layers}
	\end{figure*}
	
	\subsubsection {IoT Network Stack} \label{ IoTNetworkStack}

	The IoT network stack can be described as being built on a protocol of layers \cite{rayes_salam_2016}, as illustrated in Figure \ref{fig:iot_layers}. Each layer plays an integral role in ensuring that the automated tasks initiated at the sensor are completed at the actuator. The layers of the IoT protocol stack work together to provide a complete communication infrastructure for IoT devices, enabling them to efficiently exchange data with each other and over the internet. While the individual layers of the architecture combine to create the IoT, the authors of \cite{mrabet_belguith_alhomoud_jemai_2020} illustrate that each layer of the IoT stack represents unique challenges for cyber security researchers and opportunities for cybercriminals to exploit security and privacy vulnerabilities. This investigation will now briefly explain the security and privacy challenges of each layer of the IoT stack and its role in IoT infrastructure, demonstrating that a single solution for privacy security on IoT is challenging.

	\paragraph{Physical/Link Layer}

 The first layer, often described as the bottom layer of IoT and illustrated as the left most segment in Figure \ref{fig:iot_layers}, is the physical/link layer \cite{rayes_salam_2016}. This layer consists of sensors that gather information and detect environmental changes. It then senses other connected devices within the environment to initiate communication to relay the parameters \cite{soni_upadhyay_jain_2017}. The authors of \cite{mrabet_belguith_alhomoud_jemai_2020} recognise cyber security risks at this level as eavesdropping, cyber-physical, and RFID tracking. Of these threats, eavesdropping is the most likely to result in threats to privacy. Much like eavesdropping on a conversation, the authors of \cite{mrabet_belguith_alhomoud_jemai_2020} explain an eavesdropping attack as when a cybercriminal attempts to collect information sent from the IoT device. This type of attack can result in the loss of confidential information. Eavesdropping is an example of security considerations that can lead to the loss of privacy through IoT on 5GBN networks. 
	
	A cyber-physical attack is when a cyberattack impacts the physical environment \cite{mrabet_belguith_alhomoud_jemai_2020}. While this type of attack represents the potential for significant disruption, loss of privacy is of lower concern, and the authors of \cite{mrabet_belguith_alhomoud_jemai_2020} highlight several proposed solutions to mitigate the risks from cyber-physical attacks. Similarly, RFID tracking is when cybercriminals attempt to disable, imitate, or modify the contents of an RFID tag. While RFID tracking attacks can result in significant security incidents, such as unauthorised access to restricted areas, the authors of \cite{mrabet_belguith_alhomoud_jemai_2020} discuss solutions researchers have developed to counter these concerns.
    Low-powered devices, such as {RFID’s}, are incapable of using physical layer security and they take leverage from the protocol stack security of the networks to secure their information for communication as highlighted in \cite{kumar2023ultra, chiang2016fog}. Using the network edge for security has several benefits, such as improved cost of transmission, scalability and superior security once the data has reached the edge. However, the benefits gained by using the network edge do come with risks. Although the network can secure the data transmitted, data collected by the device still remains vulnerable to cyber-attack until the edge has secured it. As identified above, failing to secure information will allow cybercriminals to exploit this vulnerability. Unlike more powerful devices, such as mobile phones which can implement physical layer security at the transmitter and receiver, ensuring enhanced end-to-end security, low-powered IoT devices lack this capability.

    \paragraph{Network Layer}

    According to the authors of \cite{mrabet_belguith_alhomoud_jemai_2020} and \cite{vashi_ram_modi_verma_prakash_2017}, privacy vulnerabilities are also present at the next layer in the IoT stack, which is the network layer. The network layer allows communication between IoT sensors using various networks, such as Bluetooth, Wi-Fi, 5G and other network connections \cite{mahmoud_yousuf_aloul_zualkernan_2015, rayes_salam_2016}. As with the physical layer above, eavesdropping remains a privacy concern at the network layer along with Man-In-the-Middle (MIM) attacks \cite{mrabet_belguith_alhomoud_jemai_2020, mahmoud_yousuf_aloul_zualkernan_2015}. A MIM attack occurs when an attacker modifies the correspondence between parties who trust the communication between themselves \cite{mallik_2019, callegati_cerroni_ramilli_2009}. As noted by the authors of \cite{nathnayak_ghoshsamaddar_2010}, a MIM attack can result in not only the modification of data but also the loss of privacy. While the authors of  \cite{mrabet_belguith_alhomoud_jemai_2020} note that several solutions are available to prevent a MIM attack, they also discuss computational power limitations present in IoT for advanced protective measures. The authors of \cite{mohamadnoor_hassan_2019} discussed this limitation earlier, particularly regarding authentication and transport encryption. Transport encryption refers to data transmission across the internet using secure encryption technology.

 \paragraph{Transport Layer}

 As explained by the authors of \cite{lin_bergmann_2016}, the transport layer of the IoT stack introduces the first step of true IoT security and privacy enhancements through data encryption. Although IoT lacks the resources for Transport Layer Security (TLS), a cryptographic protocol for providing network security, IoT employs Datagram Transport Layer Security (DTLS), which provides similar cryptographic security to TLS \cite{lin_bergmann_2016}. DTLS is also preferable in IoT applications as it has lower latency. However, despite the introduction of encryption, the authors of \cite{yang_wu_yin_li_zhao_2017} and \cite{assiri_almagwashi_2018} explain that the transport layer is still vulnerable to cyber-attacks, which can lead to loss of privacy. While encryption offers safeguards against cybercriminal activity at this level, unencrypted connections are susceptible to MIM attacks, leading to direct loss of privacy and other cyber security concerns \cite{mallik_2019}. Additionally, the authors of \cite{mrabet_belguith_alhomoud_jemai_2020} identify resource exhaustion, flooding, replay, and amplification attacks commonly orchestrated by cybercriminals against the transport layer. Of the types of attack recognised by the authors of \cite{mrabet_belguith_alhomoud_jemai_2020}, a replay attack is the most likely to lead to loss of privacy. A replay attack is a type of MIM attack. In this attack, a cybercriminal eavesdrops on secure network communications, intercepts them, and then fraudulently delays or re-sends the message to misdirect the receiver.

 \paragraph{Application Protocols and Application Services Layer}

 The next layer of the IoT infrastructure stack that will be examined for privacy and security concerns is the application layer. The authors of \cite{saritha_sarasvathi_2017} describe the application layer as the top layer of the Transmission Control Protocol/Internet Protocol (TCP/IP) stack. However, unlike the TCP/IP stack, the authors of  \cite{rayes_salam_2016} describe this layer of IoT infrastructure as being split into two layers: Application Protocols and Application Services. In IoT, the application layer connects the device and the network with which it will communicate \cite{donta_srirama_amgoth_raoannavarapu_2021}.
	As the authors of \cite{donta_srirama_amgoth_raoannavarapu_2021} note in their research, the growth of applications at this layer in the IoT stack is not only generating new opportunities for enhanced security but is also creating new cyber security challenges and opportunities for cybercriminals. While the authors propose using ML as a potential solution to many of the security issues faced by developing applications, ML also has its own challenges. The authors of \cite{suomalainen_juhola_shahabuddin_mammela_ahmad_2020} address the concerns of blindly using ML in wireless communication, which can have disastrous results and can lead to the exposure of critical network infrastructure to cybercriminals and will be discussed in detail in Section \ref{MachineLearnign}.

In summary, IoT characteristics contribute significantly to the security concerns of the technology. From the low processing power that limits security enhancements to the network stack, personal IoT devices exhibit several unique characteristics that contribute to the vulnerabilities of the technology. While the network stack has the potential to enhance security by using contemporary digital technologies, each layer possesses unique challenges to protecting the security of data collected by personal IoT devices. 
Despite efforts to add encryption at the most vulnerable layer by the authors of \cite{khatib_ahmed_kamaleldin_abdelghany_mostafa_2018} and \cite{tsai_leu_you_chang_hu_park_2019}, encryption is yet to become an embedded feature of IoT devices. Although encryption will degrade battery life and processing performance, improved battery life and faster microprocessors will reduce encryption costs, enhancing security and privacy, which vastly outweigh the negatives.
Although researchers have suggested solutions to many of the cyber security concerns raised by the characteristics of IoT, more research is required to address the security concerns and vulnerabilities identified in this paper, particularly at the physical layer. 

\begin{figure*}[t]
		\centering
		\includegraphics[width=0.85\linewidth]{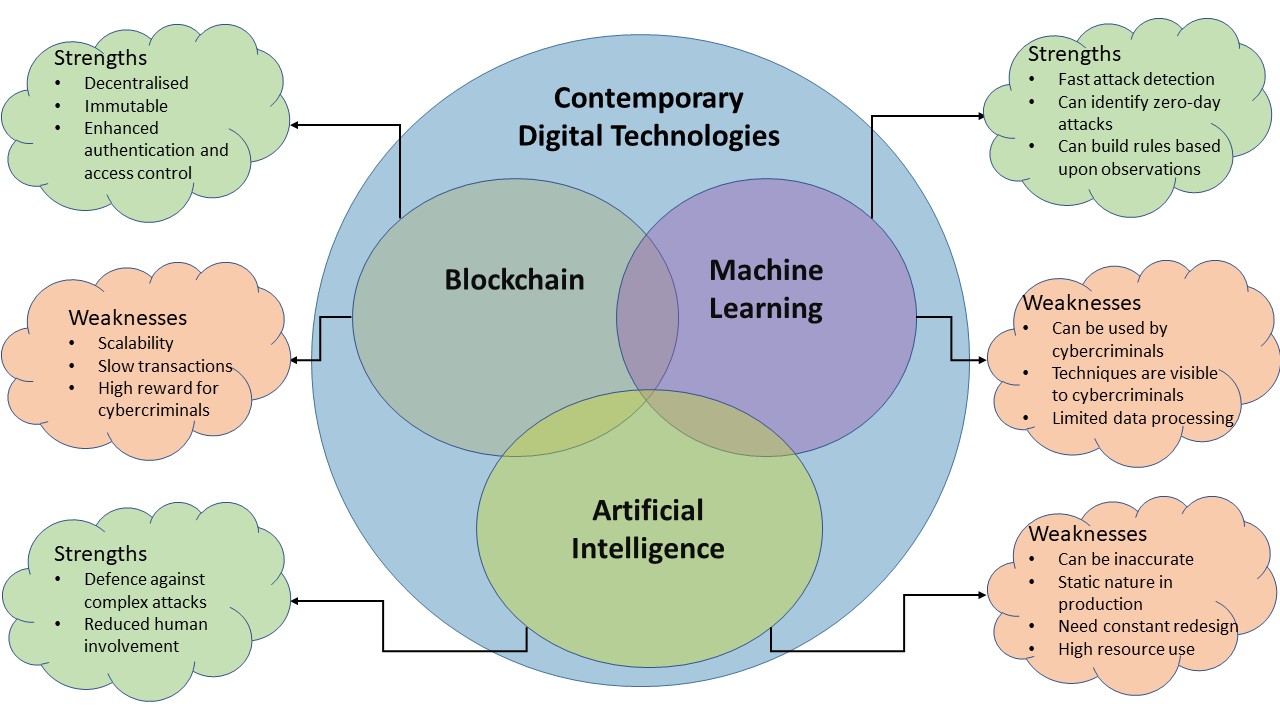}
		\caption{Contemporary digital technologies strengths and weaknesses}
		\label{fig:security_and_privacy_issues_in_the_6G_network}
	\end{figure*}

 \subsection{Contemporary Digital Technologies} \label{ContemporaryDigitalTechnologies}
 Researchers have identified five main areas of concern to privacy and security on futuristic networks  \cite{wang_zhu_zhang_zhang_yu_zhou_2020}.
 They are authentication, access control, malicious behaviour, encryption, and communication. With the increased usage of personal IoT devices and a progression to evolving futuristic networks, researchers have started to investigate the use of contemporary digital technologies such as AI, ML and blockchain to enhance the security of these low-powered devices \cite{taylor_dargahi_dehghantanha_parizi_choo_2019, alotaibi_2019, ma_2021, b.s._s._kashyap_d.n._2019}. This paper will now investigate the three most researched technologies, AI, ML and blockchain, to determine the strengths, weaknesses, and suitability of each technology in enhancing the security and privacy of low-powered personal IoT devices on 5GBN, as illustrated in Figure \ref{fig:security_and_privacy_issues_in_the_6G_network}.

 \paragraph{Machine Learning}\label{MachineLearnign}
	ML is a technology often recommended alongside AI as a solution for enhanced cyber security for IoT devices\cite{unsal_ustun_hussain_onen_2021, he_miari_makrani_aliasgari_homayoun_sayadi_2021}, often due to its placement on the network edge \cite{bourechak2023confluence}. Although ML offers significant benefits in enhancing security and privacy, one of the major disadvantages of the technology is the ability of cybercriminals to also use it as a tool to circumvent protective measures \cite{brundage_avin_clark_toner_eckersley_garfinkel_dafoe_scharre_zeitzoff_filar_etal._2018}. Due to its tendency to be resource intensive, ML is often implemented later in the IoT stack, resulting in the physical layer remaining vulnerable to exploitation. However, researchers are exploring ways to optimise resources using federated learning for use in IoT \cite{da2023resource}. To understand the advantages and disadvantages of ML as a tool to enhance cybersecurity, it is necessary to explain how the technology works.

ML is used in many daily settings to enhance security and privacy online. ML is the process of building algorithms for analysing and predicting results through data and statistics \cite{wazid_das_chamola_park_2022, mitchell_1997}. For ML to be effective as a tool to fight cybercrime, it relies on a set of rules to work on as a guide. These rules guide ML in making decisions in real-time to enhance security and privacy \cite{mitchell_1997}. A typical daily use case of ML that many consumers use daily is the avoidance of spam emails. In the email use case, the email provider or a user inputs a set of rules, such as email addresses, subject lines, keywords, IP addresses or hostnames and the spam filter will use those rules in an algorithm to filter spam \cite{dada_bassi_chiroma_abdulhamid_adetunmbi_ajibuwa_2019}. While IoT should not be considered a comparable technology to email, the application of ML in the setting of IoT is fundamentally similar. Rules can be created to create a set of filters that can instantaneously enhance the security of the devices that use them without the need to be explicitly programmed \cite{hussain_hussain_hassan_hossain_2020}. By monitoring how the data from the devices is accessed, ML can act and block suspicious activity \cite{agarwal_sharma_alshehri_mohamed_alfarraj_2021}. A significant benefit of this approach is that zero-day or new exploits can be quickly identified, and rules deployed to counter the attack \cite{he_miari_makrani_aliasgari_homayoun_sayadi_2021, wazid_das_chamola_park_2022}. However, as the author of \cite{li_2018} notes, ML has an over-reliance on feature extraction. This can result in new threats being left undiscovered as no feature rule set was devised to detect them \cite{li_2018}. Although the rapid deployment of new features in ML can promptly remedy an attack vulnerability, the very nature of ML as a technology means that it can also be used as a tool to discover and circumvent security systems \cite{brundage_avin_clark_toner_eckersley_garfinkel_dafoe_scharre_zeitzoff_filar_etal._2018}.

Although ML is a valuable tool for enhancing cyber security for IoT devices, it possesses several key characteristics that expose its weaknesses and limitations as a robust solution against cybercrime. Its principle weakness is its ability to be used as a cybercriminal tool to exploit vulnerabilities in systems \cite{brundage_avin_clark_toner_eckersley_garfinkel_dafoe_scharre_zeitzoff_filar_etal._2018}. A cybercriminal can use ML to continually probe a system and gain knowledge about the defences in place. The knowledge built during these probing attacks can be used to form a successful attack against a system. However, despite the risks of cybercriminal misuse of the technology, as the authors of \cite{wazid_das_chamola_park_2022} note in their research, ML remains a significant tool for cyber security protection.

\paragraph{Artificial Intelligence}\label{ArtificialIntelligence}
%\textbf{\textit{Artificial Intelligence}}: 
One of the leading technologies in cyber security defence is the use of AI. According to a survey by Information Technology (IT) consulting firm, Capgemini Research Institute, 69\% of the respondents from a survey of 850 senior IT executives stated that the use of AI will enable them to effectively respond to cyberattacks, while 61\% believe AI is essential for identifying threats \cite{tolido_thieullent_vanderlinden_frank_delabarre_buvat_theisler_cherian_khemka, lazic_2019}. These figures indicate a significant reliance on AI as a tool to enhance cyber security. However, a deep-seated trust in AI as a robust solution to cyber security concerns may be misplaced. While AI is proving to be a reliable mainstay in cyber security, it has encountered limitations that reduce its effectiveness not only as a tool for security enhancement but also as a tool used by cybercriminals \cite{brundage_avin_clark_toner_eckersley_garfinkel_dafoe_scharre_zeitzoff_filar_etal._2018}.

To understand the relevance of AI as a cyber  security tool, it is important to learn what AI is and how it is used in cyber security. A significant advantage of AI as a tool for cyber security is the ability of the technology to learn and behave independently from system administrators \cite{li_2018}. Where a human would be required to perform certain tasks and checks, an AI system is a purpose-built tool that automates the detection and decision-making processes. By using a statistically weighted matrix, also known as a neural network, AI replicates human decision-making using a systematic and rationalised approach \cite{li_2018}. The neural network is a decision matrix where AI applies deep learning. The interconnected nodes serve as weighted biases in each filtering step, where certain rules are given a higher value than others. After pre-compiling the data, it is stored in a database, which is received by the neural network. As the system learns from the information gathered from previous decisions and new data, it improves its knowledge of its task. AI creates a bias based on the collected data, which it then learns and resolves answers based on the data analysed. Although automated, the bias follows a set of rules defined in the modelling design, which may include and exclude certain observations. The data that is analysed is then subject to the design rules and returns relevant results based on those rules. While AI provides significant automation benefits, it has several key features working both in its favour and against it.

The main advantage of AI in cyber security is its ability to defend against complex attacks \cite{li_2018}. Today, cybercriminals are honing their skills and developing more complex methods of attack capable of yielding greater rewards \cite{dupont_lusthaus_2021}. This compares to attacks in the past that typically consisted of simple trojans and viruses. As the author of \cite{li_2018} notes, complex attacks require complex cyber security solutions. As technologies and systems evolve and cybercriminals begin using more complex tools, such as AI to circumvent cyber security defences, more advanced tools must be developed and deployed to meet these challenges \cite{li_2018}. Thus, AI acts as a double-edged blade. While it can be used as a tool to combat and help protect sensitive systems from cybercriminal activity, the technology itself is increasingly being used by cybercriminals to exploit systems \cite{brundage_avin_clark_toner_eckersley_garfinkel_dafoe_scharre_zeitzoff_filar_etal._2018}. As the author of \cite{li_2018} notes, sophisticated attacks utilising AI and ML require equally sophisticated tools for defence, and AI is one such tool. However, while AI is a powerful tool for cyber security, it has several features that limit its effectiveness. 

A significant limitation of AI as a cyber security tool is its accuracy \cite{kim_park_2020}. The accuracy of AI relies in a large part on the amount of resources available. A large neural network with many decisions will require significant resources \cite{zhao_li_qi_xu_2020}. As mentioned in Section \ref{SecurityPrivacyData}, IoT devices do not possess the resources to run complex computations. This means that AI must be implemented later in the IoT network stack, such as the edge \cite{bourechak2023confluence}. However, while AI can be outsourced to related systems, these systems have resource capacity limits that inhibit the high data requirements of AI \cite{zhang_ning_shi_farha_xu_xu_zhang_choo_2021}. To counter the capacity limitations, the accuracy of AI is compromised, resulting in systems that are either too stringent with their rules or inadequate in their determination of malicious activity \cite{zhang_ning_shi_farha_xu_xu_zhang_choo_2021}. If a system is too strict, it may inhibit legitimate use, especially with highly automated systems that incorporate IoT, and as a result, the rules are often relaxed \cite{truong_zelinka_2020}. This results in systems that may be vulnerable to exploitation. 

Another limitation of AI as a suitable candidate for IoT security is the nature in which it operates in a production environment. Once an AI model is trained, it is often not re-trained in service, resulting in a static environment that could be vulnerable to cyberattack \cite{truong_zelinka_2020, zhang_ning_shi_farha_xu_xu_zhang_choo_2021}. There are several reasons that AI remains static in production usage. As previously mentioned, hardware resource availability limits the processing capacity of AI training. In a production environment, new data will be arriving continually, forcing the AI to train from newly available data, increasing hardware usage. An additional risk is raised if the AI system learns something new that makes it less effective \cite{zhang_ning_shi_farha_xu_xu_zhang_choo_2021}. Human-centric issues also have a hand to play. An error arising from an operator incorrectly adjusting data can additionally result in similar outcomes. To ensure that software meets specification requirements during use, AI often remains in a static state and will be updated to counter new exploits when they arise \cite{zhang_ning_shi_farha_xu_xu_zhang_choo_2021}. This can result in a vulnerability gap before the software is updated.

	\paragraph{Blockchain}\label{Blockchain}
	Blockchain is a novel technology that has gained popularity among researchers to enhance authentication, access control and communication. Blockchain came to prominence with the introduction of Bitcoin in 2008 \cite{ chowdhury_colman_kabir_han_sarda_2018}. According to the authors of \cite{nguyen_ding_pathirana_seneviratne_li_niyato_dobre_poor_2021}, a blockchain is a decentralised, immutable, and transparent database that operates on a ledger-based system and enhances authentication, access control and communication \cite{wang_zhu_zhang_zhang_yu_zhou_2020}. While blockchain provides significant security and privacy enhancements on 5GBN, it exhibits several traits in its current form that make it unsuitable for IoT applications on futuristic networks. Principle among these traits are slow transaction processing speed of data on the blockchain \cite{ omar_jayaraman_salah_yaqoob_ellahham_2020, patel_khatiwala_shah_choksi_2020, lemieux_2016} and scalability \cite{ khan_jung_hashmani_2021, chowdhury_colman_kabir_han_sarda_2018}, both of which will be examined below in addition to security concerns of the technology. Although research in these specific areas is ongoing, their limitations highlight the current unsuitability of blockchain technology as a reliable candidate for securing IoT data on emerging 5GBN.

When addressing the privacy and security requirements of IoT on futuristic networks a factor that needs to be addressed is the speed of the data transmission and its scalability. While being considered as a secure system, blockchain itself suffers from slow performance \cite{kim_kwon_cho_2018}, with Bitcoin transactions being limited to only seven transactions per second \cite{chauhan_malviya_verma_mor_2018}. \textit{Slow processing} of IoT transactions through blockchain technology on such networks that require real-time data is counterproductive to the needs of moving to futuristic networks that will be discussed in Section \ref{Beyond5G}. The slow processing time can be attributed to the nature of blockchain. Much like a traditional database, the primary function of blockchain is to store data, making it a type of database \cite{tasatanattakool_techapanupreeda_2018}. As a database grows the speed of the response time to the queries from the database decreases \cite{fisher_2011}. As explained by the authors of \cite{fisher_2011}, a database containing millions of rows cannot complete the requested query in real-time. Herein lies a fundamental flaw with the reliance of blockchain as a tool to secure IoT data on futuristic networks. Compared to traditional databases that are centralised and controlled by individuals, organisations, or groups, blockchain is a decentralised database that transfers the control and decision making to a distributed network \cite{tasatanattakool_techapanupreeda_2018}. However, as demonstrated by the authors of \cite{arif_khan_rehman_kabir_imran_2020}, \cite{newell_mamun_rehman_islam_2022} and \cite{shabandri_maheshwari_2019}, research is already underway to improve the speed and efficiency of blockchain transactions. 

The authors of \cite{arif_khan_rehman_kabir_imran_2020} identified that using blockchain as a security method for smart home applications brought with it concerns that it may not fulfil the demand for smart home security. The cause of the concern is the scalability issues arising from the nature of blockchain, identified earlier in this paper. The authors noted that the scalability concerns are due to the ability of anyone having the ability to join the network. This results in increased network costs. The solution proposed by the authors is the implementation of a  consortium blockchain. A consortium blockchain is a combination of public and private blockchain that work together to share information to improve existing workflows \cite{dib2018consortium}. By utilising this approach, the user’s performance as a node in the blockchain process is eliminated. The result is reduced network costs and the ability to scale.

In the research by the authors of \cite{newell_mamun_rehman_islam_2022}, they were able to improve the speed of the transactions by replacing the consensus algorithm, known as Proof-of-Work, with a more efficient algorithm they called Proof-of-Enough-Work (PoEW). Their method increased the efficiency of computing resources to process a block resulting in a significant increase in the transaction processing rate. When compared to the bank transaction processing rates of both Swift and Visa, the PoEW method surpassed both. This indicates a vast improvement in the transaction rate. Although physical drive read and write limits will still inhibit scalability, this method vastly improves the blockchain scaling ability.

The authors of \cite{shabandri_maheshwari_2019} were also able to improve the blockchain transaction processing time by improving the efficiency of the computing resources. In their experiments, the authors used  Internet of Things Application (IOTA), which is a decentralized, open-source cryptocurrency specifically designed for IoT. In their approach, they remove two key inhibitors that reduce the suitability of blockchain as a security solution for IoT. These are transaction fees and the concept of mining which require large amounts of processing power  \cite{shabandri_maheshwari_2019}. In addition to this, the key concept is the application of a guided acyclic graph for transaction storage which greatly improves the efficiency of blockchain. A positive outcome of this approach is that the transaction speeds increase as more devices join the network. There is, however, a limitation to this approach. Its effectiveness is highly dependent on it popularity and uptake. The reliance on more devices to increase processing speed is a metric which cannot be guaranteed in a commercial environment.

Despite the advances made in recent studies \cite{arif_khan_rehman_kabir_imran_2020}, \cite{newell_mamun_rehman_islam_2022} and \cite{shabandri_maheshwari_2019} and the benefits of blockchain as a cyber security tool, the popularity of blockchain has made it an enticing target for cybercriminal activity \cite{wang_wang_hu_2019}. While blockchain exhibits many favourable characteristics for enhanced security and privacy, it does possess \textit{security vulnerabilities} that can result in privacy being compromised. A well-known vulnerability of blockchain is a fifty-one percent attack \cite{ye_li_cai_gu_fukuda_2018}. Such an attack occurs when a malicious attacker gains more than fifty percent of the blockchain ledger. As highlighted in \cite{ye_li_cai_gu_fukuda_2018}, this type of attack allows the perpetrator to modify the blockchain transactions. Although researchers have proposed solutions to this type of attack \cite{ye_li_cai_gu_fukuda_2018}, additional research must be undertaken to ensure its feasibility for IoT devices on futuristic networks. A fifty-one percent attack is but one of the known vulnerabilities of blockchain. Another common vulnerability is the forking attack where an attacker attempts to launch an alternative chain to the most trusted chain in the blockchain \cite{wang_wang_hu_2019}. This attack, if successful, can lead to fraudulent transactions in the blockchain. A forking attack is particularly hazardous for users of personal IoT devices because if the forking attack is a hard attack, it is not reversible \cite{ramos_pianese_leach_oliveras_2021}. Further security concerns of blockchain have been highlighted by the authors of \cite{ramezanpour2022security} who identified that blockchain on a wireless network that shares database access with the network is prone to a range of security threats that can result in cybercriminals emulating a hidden node. This can result in a MIM attack or spectrum hijacking.

In summary, while each of the contemporary digital technologies discussed can be used to enhance the security of personal IoT devices on 5GBN, they each exhibit limitations that can result in gaps in security defences. As discussed earlier, individual IoT devices do not possess the computational power needed by these technologies, and as a result, they are implemented later in the network stack. This leaves the physical layer of the network stack where the personal IoT device sits vulnerable to direct exploitation. Despite this limitation, the contemporary digital technologies discussed have an important role to play in IoT security. While the authors in \cite{newell_mamun_rehman_islam_2022} and \cite{shabandri_maheshwari_2019} offer novel solutions to the scaling issues of blockchain identified in this paper, their commercial application may be limited. In another study, \cite{han_gramoli_xu_2018} authors have identified that enterprises have demonstrated reluctance to implement proof-of-work concepts as they offer probabilistic guarantees and can be subject to double spending. It is foreseeable that a future application of blockchain in IoT will incorporate a consortium blockchain approach discussed by \cite{arif_khan_rehman_kabir_imran_2020} alongside methods designed by \cite{newell_mamun_rehman_islam_2022} and \cite{shabandri_maheshwari_2019} to reduce the cost of the blockchain transaction further and increase its scalability and suitability to a rapidly growing IoT industry. It is, therefore, essential that additional research be undertaken in this area. While these technologies continue to evolve, an important factor in the security of personal IoT devices is the regulation of the devices and communication of the data. Although contemporary digital technologies can be utilised to enhance security of the devices, failure to develop workable standards that can be widely implemented can render such contemporary digital technologies worthless if the device is not secure by design.

	\subsection{Human-Centric Issues Regarding IoT Privacy and Security} \label{HumanCentricIssuesRegardingIoTPrivacySecurity}
 
	Today, IoT devices are present in a multitude of settings \cite{rayes_salam_2016}. Government departments such as police and customs rely on Internet Protocol (IP) cameras to detect biometric features for crime prevention and improve efficiency in immigration processing \cite{yimyam_kocento_ketcham_2018}. Federal aviation safety regulators use GPS sensors to track aircraft movements to enhance safety and improve response time during emergencies \cite{matthews_nielsen_schade_chan_kiniry_2014}. Primary production has implemented IoT to improve agricultural efficiency with automated harvesting, irrigation, and the sowing of crops \cite{lee_hwang_yoe_2013}. The automotive industry has adopted innovative technology for crash avoidance, smart parking, and autonomous navigation \cite{krasniqi_hajrizi_2016}. 
	
	The improvements to efficiency and quality of life demonstrated above are some of the key benefits offered by IoT as additionally highlighted in the study undertaken by \cite{seliem_elgazzar_khalil_2018}. However, while these benefits are a driving factor in the adoption of IoT, the researchers have identified a sudden increase in new users and an increased number of devices acting as contributing factors due to poor cyber security protocols of IoT devices \cite{tawalbeh_muheidat_tawalbeh_quwaider_2020}. Additionally, the authors of \cite{seliem_elgazzar_khalil_2018} and \cite{tawalbeh_muheidat_tawalbeh_quwaider_2020} identify privacy as a significant cyber security challenge for IoT.
	
	While beneficial for the user, IoT efficiency and automation present significant cyber security risks. As identified by the authors of \cite{seliem_elgazzar_khalil_2018} and \cite{tawalbeh_muheidat_tawalbeh_quwaider_2020}, privacy is currently a significant challenge for IoT. The authors did not identify a principal cause of the privacy challenges; instead, they discovered that the privacy challenges originate from many sources. While neither research identified one single cause, common themes are the lack of a standard security scheme and reliance on the end-user to protect their devices with strong passwords and regular software update maintenance \cite{tawalbeh_muheidat_tawalbeh_quwaider_2020, seliem_elgazzar_khalil_2018}. The authors of \cite{tawalbeh_muheidat_tawalbeh_quwaider_2020} further identify the sudden increase of new users and an increased number of devices as contributing factors to poor cyber security protocols of IoT devices. The factors identified are an abundance of weak password policies and a failure to ensure device software is maintained to a current secure version. 
	
	As stated, reliance on the end-user to conduct software updates for IoT devices increases the risk of loss of privacy in addition to device and network vulnerability exploitation \cite{vaniea_rashidi_2016}. While not apparent to users, vulnerabilities can exist in systems and may remain undiscovered for extended periods. Vulnerability discovery often falls upon two groups of people, ethical hackers, and cybercriminals. Over time, ethical hackers will seek to discover vulnerabilities in systems. An ethical hacker is a professional that companies employ to test and detect vulnerabilities in software and systems. Successful ethical hacking allows companies to patch vulnerabilities before cybercriminals exploit them \cite{patil_jangra_bhale_raina_kulkarni_2017}. Patching is the process of releasing changes that fix, alter, repair, or improve security vulnerabilities or other bugs in software \cite{dadzie_2005}. Ethical hacking offers developers, manufacturers, and managers of the device an opportunity to update to a secure version before the vulnerabilities are discovered and exploited by cybercriminals. Occasionally, cybercriminals will discover vulnerabilities prior to ethical hackers or security researchers and begin exploiting the vulnerabilities for malicious gain \cite{rennie_shore_2007}. Malicious exploitation creates an urgency for the software to be patched, leaving end-users vulnerable to known and exploitable security vulnerabilities. 
	
	As explained by the authors of \cite{tawalbeh_muheidat_tawalbeh_quwaider_2020} and supported by the authors of \cite{coppens_desutter_debosschere_2013} in their research paper, the reliance on the end-user to update their devices is unreliable and can expose the users' device and network to cyber security harm if the update has been delayed or missed. To counter this reliance, some researchers have suggested the use of automated software updating as a means of protecting the software \cite{coppens_desutter_debosschere_2013}. However, the authors of \cite{wash_rader_vaniea_rizor_2014} identified that automated software updates can lead to undesirable outcomes. The use of outdated software, which may contain security vulnerabilities, is exacerbated by end-user's weak password policies that may be exploited by cybercriminals \cite{tawalbeh_muheidat_tawalbeh_quwaider_2020}.
	
	To summarise, the human-centric challenges identified in this section highlight significant barriers to securing IoT device confidentiality on 5GBN. However, the challenges of IoT confidentiality are not limited to human-centric issues. The challenges identified in this section by the authors of \cite{seliem_elgazzar_khalil_2018} and \cite{tawalbeh_muheidat_tawalbeh_quwaider_2020} go beyond the individual devices, with exploits possible across several layers of the IoT architecture  \cite{mrabet_belguith_alhomoud_jemai_2020}, which was discussed in detail in Section \ref{ IoTNetworkStack}.
	
	\subsection{Security and Privacy of Data Collected on IoT Devices } \label{SecurityPrivacyData}

 While the human-centric challenges to privacy and security of IoT devices offers unique problems for researchers to solve, the problems identified in the previous section can be exacerbated by security flaws in IoT devices and misuse of the data collected by them. One such example of misuse is the oversight of how the data collected may be used. The potential of privacy exploitation originating from oversights in the use of technology was illustrated by the authors of \cite{moncrieff_venkatesh_west_2009} in 2009. Their research demonstrated early examples of the unintended use of personal data originating from public surveillance and the impacts on privacy. 
	In their research, they discuss Google's Street View (GSV) and the approach used by Google to address privacy concerns from the publicly available collection of data. The solution deployed by Google to preserve privacy was the removal of data \cite{moncrieff_venkatesh_west_2009}. While in the case of GSV the solution preserves privacy by removing identifiable data, such as faces and number plates from vehicles, the application of this policy in IoT devices is challenging due to computational processing power limitations \cite{zeadally_tsikerdekis_2019} discussed in detail in Section \ref{IoTCharacteristics}. 
 
 Another approach to protecting privacy investigated by the authors of \cite{moncrieff_venkatesh_west_2009} is the development of a privacy policy as a mechanism for privacy preservation. However, as identified by the authors of \cite{subahi_theodorakopoulos_2018}, the development of a privacy policy does not guarantee that it will be fully implemented. The authors of \cite{subahi_theodorakopoulos_2018} illustrate this fundamental failure with the privacy policy solution discussed by the authors of \cite{moncrieff_venkatesh_west_2009}. The presence of a privacy policy does not guarantee privacy protection, with the authors of \cite{subahi_theodorakopoulos_2018} identifying that half of all privacy policies do not adhere to their stated policies. Further to that, the collectors of confidential data rely on self-regulation for privacy safeguarding, for example, as the authors of \cite{fuchs_2012} explain with the privacy management by Facebook.
	
	Since the inception of the internet, tools have been developed that have the potential to expose an individual’s personal data. As is often the case, legislative tools designed to protect the public from exploitation often lag advances in technology. However, as demonstrated in the European Union (EU), regulators are beginning to catch up with technology with the introduction of the General Data Protection Regulations (GDPR) legislation \cite{voigt_vondembussche_2017}. The role of the GDPR, otherwise known as the Cookie Consent Law, is to protect internet users’ privacy online \cite{hoofnagle_vandersloot_borgesius_2019}. Websites that operate and are accessible to EU citizens must comply with the stringent requirements of the GDPR that guarantee a right to anonymity with harsh financial penalties in place for companies who fail to abide by the laws \cite{ hoofnagle_vandersloot_borgesius_2019}. While the introduction of GDPR laws to enhance privacy online do introduce levels of certainty to personal data protection for EU citizens, these laws are not applicable worldwide, and as identified by the authors in \cite{ystgaard2023review}, the GDPR framework has also been criticised for unclear responsibilities in some complex scenarios and offers only limited protections in others. Consequently, IoT on 5GBN introduces new paradigms of uncertainty for privacy online. 
	
	Cybercriminals will often attempt to exploit weaknesses in a system to gain access to confidential information \cite{choo_2011}. While systems are often prepared for cyberattacks or have mechanisms in place to limit the impact of a cyberattack, private information can be revealed through not only a sophisticated intrusion of systems, but also through access to freely available data which has been left unsecured either by design or by accident. Examples of such data are geolocation data, flight tracking information and daily routines. While much of this type of data can be considered of little value, there have already been incidents where this data has been exploited. When exploited, it can lead to loss of privacy and security with real world consequences. A notable example occurred in 2018 when the Strava fitness tracking app unintentionally exposed sensitive US military bases in the Middle East. This was achieved by accessing the publicly available global heatmap from the Strava fitness app and analysing the GPS data \cite{bogle_2018}. This discovery prompted an investigation by the US military into the incident \cite{bogle_2018}. This data exploitation was later expanded by Norwegian broadcasting company, NRK, who uploaded the Strava data into third party software which allowed them to identify the profiles of individual European soldiers who used the Strava fitness app \cite{rettberg_2020}.

While the Strava incident appears to be isolated, other examples of misuse of data from IoT devices exist. In January of 2022, a Twitter account that tracks the flights of entrepreneur Elon Musk’s personal jet gained attention online \cite{neate_2022, vigdor_2022}. The account used publicly available data transmitted from the planes transponders that records the flights’ location. It was collected through a service called ADS-B Exchange which collects unfiltered flight data \cite{vigdor_2022}. While the data transmitted by the transponders is mandated by the United States Federal Aviation Administration (FAA) \cite{limitingaircraftdatadisplayedfederalaviationadministration_2018}, this incident demonstrates how an oversight in how data is managed can be misused. Although it would appear to be of little value, this information can be misused to cause personal harm to the person being tracked. While there is no way to validate that the flight being tracked belongs to Mr Musk, this incident serves as an example of how public data can be used in ways for which it was never intended.

With the introduction of smart speakers, smart toys, smart cameras and smart homes, the invasion of personal privacy and security extends to not only the actions people perform with a device directly but also their daily activities and conversation, particularly in the case of smart speakers and IP cameras which are always on \cite{lutz_newlands_2021}. Further to that, the data collected via such devices may not always be wilfully granted by the individuals, or secured \cite{streiff_kenny_das_leeth_camp_2018}. This is of particular concern for smart toys due to children being the target market. While it would be assumed the interactions of a child with their toy would not illicit nefarious activity, the Fisher Price Bear smart toy is an example of how a toy can be targeted by cybercriminals. The Fisher Price Bear was a smart toy that allowed interactions through a variety of communication technologies. While many of the sensors were invasive of privacy alone, researchers discovered that the devices were insecure, potentially allowing cybercriminals root access to the smart toy with full access control to the nose camera and other sensors on the toy \cite{streiff_kenny_das_leeth_camp_2018}. This incident may be attributed to the nature of the IoT device, however, well known devices common in many households have been known to harbour security exploits \cite{jia_xiao_yu_cheng_liang_wan_2018, godwin_glendenning_gagneja_2019}. 

Smart speakers, such as Google Home and Amazon Echo have contained exploitable vulnerabilities. Attacks on Google Home using a smart phone have been shown to demonstrate effectiveness in infiltrating the target device. These attacks commonly targeted the authentication and communication process \cite{jia_xiao_yu_cheng_liang_wan_2018}, demonstrating one of the most well-known vulnerabilities of IoT devices. Another example of smart home exploitation was the first-generation Amazon Echo which contained Bluetooth, Blueborne and internal Wi- Fi network vulnerabilities \cite{godwin_glendenning_gagneja_2019}. Although these vulnerabilities have been patched, and the authors note that it is not yet possible to exploit the current versions of the Echo device, the insecure devices offer a treasure-trove of personal data to cybercriminals.

	In summary, the security and privacy of data collected on IoT devices has many challenges. While oversights into how the data may be used in ways it was never intended may seem harmless, there are real-world consequences due to the mishandling of sensitive information. Though the number of incidents is sparse, they are serious, nonetheless. With an increase in the number of devices that use intrusive technology such as IP cameras and microphones and few safeguards to ensure the devices are secure, users of personal IoT devices are exposed to potential security incidents that arise. It is therefore important that personal IoT devices follow a set of minimum security standards on which to operate on.

\subsection{IoT Communication Security Standards} \label{IotSecurityStandards}

	When addressing IoT security standards they typically fall under the jurisdiction of relevant regulatory authorities in each country. In the United States of America (USA) the relevant authority is the Federal Communications Commission (FCC) who work in conjunction with the Federal Trade Commission (FTC), which is the body responsible for protecting American consumers. The Office of Communications (Ofcom) is the relevant authority responsible for IoT communication standards in the United Kingdom (UK) and in Australia the governing body is the Australian Communication and Media Authority (ACMA). While industry bodies such as IEEE Standards Association (IEEE-SA) are developing a standards initiative, there is also a considerable effort in open-source groups to develop standardisations \cite{saha_mandal_sinha_2017}. Although the open-source initiative will provide valuable contributions to enhancing personal IoT device privacy and security, the governing bodies remain the deciding factor in the implementation of any rules or standards. This paper will now briefly discuss the challenges of applying standards internationally and the move towards a unified approach to device security.

	\subsubsection{IoT Standards Challenges}
	A particular challenge in securing IoT devices on 5GBN is compliance with different standards across jurisdictions for devices that can have a global reach \cite{pal_hitchens_rabehaja_mukhopadhyay_2020, sachdev_2020}. 
 While the authors investigate the introduction of regulations to enhance 5G and beyond network security \cite{salahdine2023security}, previous research has illustrated several challenges of this approach. As identified by \cite{pal_hitchens_rabehaja_mukhopadhyay_2020}, \cite{sachdev_2020} and \cite{karie_sahri_yang_valli_kebande_2021}, ensuring that the data that IoT devices collect and transmit meets all regulatory requirements is challenging. This point is illustrated further in the research undertaken in \cite{sachdev_2020} which highlights that it is often not enough to ensure that a device meets the strictest of global requirements as doing so does not guarantee that it will meet all the requirements of other jurisdictions. To address this limitation, the author recommends that a complete international compliance review is conducted. In addition to this, the author in \cite{sachdev_2020} additionally notes that different legal provisions often reference international standards. As IoT is a new and rapidly evolving development, a set of international standards which address privacy and security has not been fully developed yet. However, in October 2019, Australia, Canada, New Zealand, the UK, and the US governments issued a statement of intent for IoT security, paving the way to develop a uniform international approach to IoT security and privacy \cite{statementofintentregardingthesecurityoftheinternetofthings_2019}. The statement agreed to by the five governing bodies includes a commitment to collaborate with the relevant standards bodies and industry to give better protection to consumers. This will be achieved by recommending that devices should be secure by design and educating the users about safeguards associated with the security of IoT devices. Though a statement of intent has been developed between like-minded governments, it does not solve the challenges of securing IoT on 5GBN. As IoT is a global network of connected sensors and actuators that communicate autonomously, as explained in Section \ref{Introduction} by \cite{atzori_iera_morabito_2010}, it requires a complete global solution to be truly secure. Failure to apply a globally agreed-upon set of standards for IoT devices on 5GBN will contribute to inconsistencies and gaps to solutions \cite{labib_brust_danoy_bouvry_2019}.

	\subsubsection{IoT Regulation in the Australian Context}
	When assessing the security regulations of personal IoT devices from an Australian perspective, it has previously been identified that the regulation of IoT in Australia could be more robust \cite{harkin_mann_warren_2022}. An insipid set of regulations can leave consumers vulnerable to security, privacy, and consent dangers \cite{manwaring_clarke_2020}. However, work in this field is slowly progressing with the Australian government publishing in 2020 a voluntary Code of Practice: Securing the Internet of Things for Consumers \cite{securingtheinternetofthingsforconsumers_2020} that followed an inquiry into digital platforms by the Australian Competition and Consumer Commissioner (ACCC) in 2019. While the voluntary code of practice is a step forward, standardisation of IoT is necessary for the robust regulation of the industry. As it currently stands, standards throughout Australia address the overall electrical safety of the device and not the security and privacy aspects \cite{harkin_mann_warren_2022}. This, however, is changing, with recommendations from the ACCC following their 2019 inquiry paving the way for a suite of reforms that include enhanced protection for consumers under the privacy act \cite{harkin_mann_warren_2022}. It is imperative to note that various potential risks to users' privacy, autonomy, and data security from IoT devices were considered by the ACCC. In the research by the authors of \cite{harkin_mann_warren_2022}, they noted that if implemented, the recommendations would result in several benefits for consumers, with an implementation of standards to protect consumers from unfair or anticompetitive trading practices among them. Although \cite{harkin_mann_warren_2022} notes that the standards will add protections against unfair and anticompetitive practices, they fall short of applying consistency to cyber security and privacy standards for personal IoT devices using wireless communications across radiofrequency spectrums, and as a result, as identified earlier by the authors of \cite{labib_brust_danoy_bouvry_2019}, will only serve to contribute to inconsistencies and gaps in existing solutions to these problems.

	In Australia, the ACMA is responsible for Australia's radiofrequency spectrum \cite{internetofthingpaper_2020}, which encompasses not only communication but also device and network standards. According to ACMA, all IoT devices must comply with existing standards \cite{internetofthingpaper_2020}, however, those devices which connect over telecommunication networks must also comply with the telecommunication standards. While many IoT devices use non-telecommunications network connections to transmit data, the devices that do use telecommunication networks come under the Radiocommunications (Short Range Devices) Standards Act, 2014, which covers Bluetooth and Wi-Fi connections \cite{ federalregisteroflegislation_2019}. While most IoT devices fall under existing regulations, ACMA has acknowledged that the evolving landscape of IoT with new devices, connection technologies and new participants who may not have experience with ACMA regulations may introduce new challenges to the regulatory framework \cite{ internetofthingpaper_2020}. However, as identified in \cite{internetofthingpaper_2020} IoT security is a global issue affecting all countries. In Australia IoT security is the responsibility of the Australian Department of Home Affairs (ADHA).

		Although device security is typically the realm of the ADHA in Australia, the ACMA notes that there are unique IoT privacy and security concerns \cite{internetofthingpaper_2020}. As previously identified in Section \ref{HumanCentricIssuesRegardingIoTPrivacySecurity}, ACMA also identified consumer awareness of emerging technologies and threats as a particular challenge that needs to be considered. With more devices entering service for personal use and becoming more popular over time, poor consumer awareness to privacy and security threats is only becoming more challenging. Additionally, ACMA notes that devices are not always developed with privacy and security in mind \cite{internetofthingpaper_2020}. While typically lacking the computational power to enhance security as discussed in Section \ref{IoTCharacteristics}, failing to take a security-first approach amplifies the privacy and security risks introduced by consumers who are complacent or not fully aware of the risks of personal IoT devices on 5GBN present them.

		To summarise this section, IoT devices currently lack a formal set of security standards on which to operate. This absence of security standards results in the reliance on device manufacturers to develop security features. Often, many devices simply lack security features, resulting in significant risk to the user. While standards in Australia do exist, they are designed with electrical safety in mind, highlighting the disparity between the advances in IoT technology and government progress to catch up to the security threats faced by the technology. While previous research has advocated for enhanced regulation of the sector, an absence of IoT security standards inhibits the development of robust regulations. This is amplified with new participants who may not be aware of existing rules governing IoT devices, particularly in Australia. Although several countries are working towards a unified approach to developing a set of standards, for the application of standards to be meaningfully effective, the development of standards must include the involvement of more countries.

		\subsection{IoT Growth and the Need for Futuristic Technologies}\label{Beyond5G}
		In Section \ref{HistoryIOTPrivacy} this paper briefly discussed the increased growth in the adoption of IoT across a multitude of industries as well as its applications for personal use. While it is difficult to ascertain precise numbers of IoT devices, several research papers have attempted to determine the number of devices currently in use and predict the number of devices in the near future. While the authors of \cite{djenna_harous_saidouni_2021} predicted the number of connected devices to exceed 30 billion by 2025, others, such as the authors of \cite{guo_yu_zhang_li_ji_leung_2021} estimates over 60 billion devices will be in use by 2025 with the potential to reach 125 billion devices in 2030, representing a 12\% year on year growth from 2017. Although it is difficult to estimate precise numbers of IoT devices in the future, it is possible to hypothesise potential numbers by studying trends and incorporating Moore's Law which states that every two years the number of transistors in a dense circuit double \cite{rayes_salam_2016}.

    Despite discrepancies between estimates,  the need to accommodate an ever-growing influx of network-connected devices is a necessity.
		
		The growth in the use of IoT devices has already resulted in unprecedented amounts of data and the need for innovative technologies capable of processing large volumes of information at faster speeds. As identified by the authors of \cite{akhtar_khan_ullah_javed_2019}, large IoT networks are already experiencing congestion on existing fixed and wireless networks due to an overwhelming number of connected devices communicating copious amounts of data at peak times. According to the authors of \cite{bhatele_titus_thiagarajan_jain_gamblin_bremer_schulz_kale_2015}, network congestion is a leading cause of performance degradation and variability. Additionally, this network congestion can result in delays in transmission and packet losses \cite{verma2022novel}. While several research papers propose solutions to congestion control \cite{akhtar_khan_ullah_javed_2019, verma_kumar_2020, bhatele_titus_thiagarajan_jain_gamblin_bremer_schulz_kale_2015, verma2022novel}, they are supplementary short-term solutions to an ever-growing problem. Consequently, as noted by the authors of \cite{guo_yu_zhang_li_ji_leung_2021}, there is a need to develop innovative technologies capable of processing large volumes of information faster. As a result, research is now investigating the development of next-generation wireless technologies.

    According to the authors of \cite{porambage_gur_moyaosorio_livanage_ylianttila_2021}, futuristic possibilities of the next generation of wireless networks are the connected intelligence in the telecommunications networks, coupled with advanced networking and AI technologies, as mentioned by the authors of \cite{wang_zhu_zhang_zhang_yu_zhou_2020}. Additionally, the authors of \cite{nguyen_lin_cheng_hwang_lin_2021} note that the anticipated increase in speed and lower latency will enable wider use of already growing technologies such as wearable IoT devices and autonomous vehicles. The authors envisage that futuristic networks will also pave the way for Three Dimensional (3D) holographic representation of individuals at virtual meetings, mixed reality, tactile internet, and implantable devices. While many potential functions of futuristic networks discussed by researchers are independent of IoT, the vast majority incorporate or are related to the IoT industry. Expanding on this, the authors of \cite{nguyen_lin_cheng_hwang_lin_2021} illustrate that one of the key visions for 6G as a potential futuristic network is an entirely autonomous network, which is precisely how IoT operates. However, as the authors of \cite{wang_zhu_zhang_zhang_yu_zhou_2020} and \cite{nguyen_lin_cheng_hwang_lin_2021} explain, there are security and privacy concerns with a futuristic 6G network and IoT that are currently being explored by researchers that will impact the security of the visions discussed.

		In addition to the network congestion concerns on existing networks, there is a growing need for faster transmission of data with lower latency that was identified as early as 2015 \cite{hung_liau_lien_chen_2015}. The low latency and improved performance requirement of IoT, especially in an industrial setting, was later supported by the authors of \cite{derhamy_eliasson_delsing_2017} in 2017. With research continuing in improving latency and overall performance on existing 5G infrastructure \cite{siddiqi_yu_joung_2019, ahmadzadeh_parr_zhao_2021, mostarda_navarra_nobili_2020}, it is becoming increasingly clear that existing networks lack the capacity and performance requirements for communication of IoT data. The slower transmission and latency delays on existing networks will worsen due to congestion, as has been previously highlighted.

  According to the authors of \cite{mahdi_ahmad_qassim_natiq_subhi_mahmoud_2021}, the next iteration of wireless networking is expected to arrive by 2030 when the number of IoT devices is anticipated to number more than thirty billion connections. One candidate to replace an eventually ageing 5G wireless network is 6th Generation (6G) networks. The next generation of wireless network infrastructure aims to meet the increased capacity demands of wireless communication of the next decade \cite{ dang_amin_shihada_alouini_2020}. However, 6G research is in the preliminary stages. The paper by the authors of \cite{kim_2021} identified that existing 6G research has primarily focused on designing antenna systems suitable for the evolving network, implementing multiple-input and multiple-output (MIMO) communication, and the development of terahertz frequencies capable of transmitting more data at a faster rate.
While research into the next generation of wireless technology has begun with an emphasis on 6G, existing research has so far had limited scope, with the standard functions and specifications still undefined. As a result, the true potential security and privacy risks have not yet been explored \cite{wang_zhu_zhang_zhang_yu_zhou_2020, porambage_gur_moyaosorio_livanage_ylianttila_2021}.
 However, as discussed by the authors of \cite{wang_zhu_zhang_zhang_yu_zhou_2020}, emerging network technologies have many possibilities, including the application of advanced AI and ML. Although the strengths and weaknesses of both AI and ML were discussed in Section \ref{ContemporaryDigitalTechnologies}, their true future potential on 6G networks is yet to be tested.

The functionality and security features of blockchain and AI will feature prominently on futuristic networks such as 6G \cite{kim_2021}. The authors identified that for 6G to meet its full automation potential, it will be dependent on the features of AI. They state that the use of AI will support intelligent edge computing, optimize resource management, and improve user detection. However, as discussed by the authors of \cite{siriwardhana_porambage_liyanage_ylianttila_2021}, the reliance on AI to enhance the functionality of 6G will result in attacks on AI systems. As a result, the authors of \cite{siriwardhana_porambage_liyanage_ylianttila_2021} identify that attacks on AI systems targeting data collection will lead to privacy issues. However, while AI will confront significant threats, the use of AI will complement other technologies such as blockchain. The authors of \cite{kim_2021} and \cite{siriwardhana_porambage_liyanage_ylianttila_2021} illustrate that the use of AI will allow for the identification of cyber-attacks in wireless networks. Further, AI will enable the detection and suppression of attacks on blockchain, such as a 51\% attack discussed in Section \ref{ContemporaryDigitalTechnologies}, allowing for a more secure network \cite{siriwardhana_porambage_liyanage_ylianttila_2021}.

The security implications for IoT devices on 6G networks are not limited to contemporary digital technologies, with security benefits and challenges also affecting the physical layer. As discussed in Section \ref{IoTCharacteristics}, the physical layer is one of the most challenging layers of IoT to protect due to the limitations of IoT devices. Futuristic networks such as 6G promise potential benefits to device security, however, personal IoT devices will continue to be vulnerable to certain types of attacks. Despite advances in contemporary digital technologies, potential characteristics of 6G, such as Terahertz (THz) technology and Visible Light Communication (VLC) technology, will remain vulnerable to certain types of adversarial activity, such as eavesdropping attacks \cite{porambage_gur_moyaosorio_livanage_ylianttila_2021}. Although the authors discussed a method to detect some forms of eavesdropping attacks on THz technology by classifying the backscatter of the intercepted channel, they note that the method does not detect all forms of eavesdropping attack. Additionally, although the authors of \cite{porambage_gur_moyaosorio_livanage_ylianttila_2021} identify that VLC systems provide heightened security benefits over radio frequency systems, eavesdropping remains a significant threat. However, while this remains a significant threat, the authors note that ML can be utilised for anomaly detection.

In summary, as the number of devices connecting to the internet increases, there is a growing urgency for developing new architectural technologies capable of sustaining the anticipated bandwidth these devices will generate at speeds faster than they currently exist. While most research has focused on 6G and the development of data communication, the full extent of privacy and security implications still need to be explored. Although AI, ML and blockchain are discussed as solutions to a litany of security concerns on futuristic networks, the physical layer of IoT infrastructure remains vulnerable to malicious activity. It is, therefore, necessary that future research address the vulnerabilities of the physical layer to further enhance IoT security on futuristic networks.

		\section{Key Findings and Further Research} \label{FurtherResearch}

  Personal IoT devices on existing networks rely on developers and individuals to maintain security of personal information collected via such devices. Due to power limitations and the need to conserve energy, cryptographic functions are often absent from many personal IoT devices. This absence of cryptographic security leaves the personal data collected on these devices vulnerable to malicious activity. This can result in personal data, including biometric information falling into the hands of cybercriminals.
  The continued absence of encryption at the device level while battery capacity is increasing and microprocessor performance is improving raises questions about the continuation of the status quo.
  As has been shown, many devices can be infiltrated and controlled due to no or low security, placing the owners of personal IoT devices at risk of harm. Additionally, the data collected can be used to reveal private information when it is made publicly accessible. Investigating the practicality of encryption at the physical layer with battery and microprocessor improvements is encouraged.

However, while the information collected at the device level is vulnerable, contemporary digital technologies such as AI, ML and blockchain have been utilised to enhance the security of the data collected. While these technologies are often proposed as a battle tested solution to cybercriminal activity, their success in mitigating the security risk of low powered personal IoT devices are limited. 
Due to the popularity of contemporary digital technologies, cybercriminals often explore them for vulnerabilities. This places devices using these technologies at risk of criminal infiltration.
As low-powered IoT devices possess limited processing capacity, they rely on the use of contemporary digital technologies to enhance security. As a result, a failure in contemporary digital technologies can expose individual devices to cybercriminal activity.
Additionally, technologies such as blockchain, in its existing form, may struggle to keep pace with the speed of transactions of 5GBN, which will demand faster processing of data. Although research is underway to solve this problem, more work must be undertaken to make it commercially enticing. Further, these contemporary digital technologies can be used as a tool to infiltrate vulnerabilities of the devices and networks that carry the information. This can place a powerful tool in the hands of a cybercriminal. Nonetheless, given personal IoT devices lack security at the physical layer, and privacy policies are often unreliable to protect an individual, the use of contemporary digital technologies to enhance security is necessary. 

Currently, the existing standards do not protect individual confidentiality of the data originating from personal IoT devices. In many cases, current and developing standards only contribute to inconsistencies and gaps in existing solutions to securing IoT data. The development of uniform standards across IoT is a development that is required to truly enhance the security and privacy of personal IoT devices on 5GBN. While personal IoT devices are a new consumer development, the need to develop a global set of recognised standards is a necessity. Currently, IoT standards in Australia typically rely on electrical safety standards for safety and communication standards. While these standards provide some level of protection for device owners, the development of security standards for personal IoT devices is a must to enhance privacy and security for users. Although the Australian government is working with four other international governments to develop standards for device security, global consistency in the application of minimal standards is necessary. 

In summary, the key findings of this investigation on the security and privacy of personal IoT devices on 5GBN are summarised as follows:
  
\begin{itemize}

	\item In addressing RQ1, this paper finds that human-centric issues, such as weak password policies and users' failure to update devices, significantly contribute to the security vulnerabilities of personal IoT devices. Additionally, due to an absence of a defined set of security standards and an over-reliance on developers to design secure systems, personal IoT security and privacy cannot currently be guaranteed. Instead, security and privacy rely on a mixture of individuals, contemporary digital technologies and device manufacturers to implement security protections.
	
	\item With a multidimensional approach to enhancing personal IoT cybersecurity, a finding addressing RQ2 is that the characteristics of personal IoT devices contribute to the security vulnerabilities of the technology with a reliance on contemporary digital technologies to fill this security void. Regardless, the physical layer remains vulnerable, particularly to eavesdropping attacks. However, with development already underway on futuristic networks, the continued absence of encryption at the device level increases data security risks. With battery capacity and microprocessor performance increasing, it is paramount that future research is conducted to investigate the feasibility of encryption at the device level. 

    \item To enhance the security of personal IoT devices, RQ3 addresses the use of contemporary digital technologies in securing personal IoT devices. A key discovery is that these technologies can be implemented to enhance the security of IoT devices, however, an over-reliance on these technologies can deliver disastrous results.

    \item From this assessment of the use of contemporary digital technologies and the earlier identified vulnerabilities, a finding which addresses RQ4 is that the absence of IoT device security standards results in many devices lacking basic security protection.

    \item In addressing RQ4 further, existing standards are primarily concerned with electrical safety with no intention for the confidentiality of data collected on IoT devices. Therefore, although an in-principal agreement exists between five nations, global standards must be implemented to protect users' privacy and security with encryption at the device level as a security consideration.

\end{itemize}

\section {Conclusion} \label{Conclusion}
		The security and privacy of data collected by personal IoT devices on 5GBN offer many challenges for  researchers. With a reliance on a mixture of security solutions, it is currently difficult to guarantee the security of personal data collected via such devices. Although the use of contemporary digital technologies can provide significant cyber security enhancements, a growing interest in the technologies by cybercriminals and an over-reliance on them as a security solution can lead to disastrous results.
 Additionally, implementing these technologies does not address other security vulnerabilities that can lead to loss of privacy at an earlier stage in the IoT stack, such as human-centric issues and physical layer vulnerabilities. With users of devices being identified as a significant inhibitor to device security, work needs to be undertaken elsewhere to enhance security.
  Although researchers have advanced cryptographic solutions to enhance security at the device level, the implementation of cryptography into devices has not yet eventuated, and as a result, the devices remain vulnerable to exploitation. With advances in battery and IoT processing performance, research investigating the implementation of encryption at the device level is encouraged.
  This, however, may depend on the development of IoT device security standards. 
  As identified in this research, an absence of IoT security standards contributes significantly to IoT device vulnerabilities, with several incidents highlighting significant flaws in data security. Although several governments are working together towards the development of IoT standards with a secure-by-design approach, the lack of consultation with the wider global community could prove to be counterproductive in the development of secure IoT devices on futuristic networks.
  However, as the next iteration of wireless technology has not yet been fully defined, the security and privacy implications are yet to be thoroughly investigated. As the next version of wireless technology approaches, it will be imperative for future research to address human-centric issues related to IoT security and enhancing security at the device level.

\bibliographystyle{IEEEtran}
%\bibliography{Thesisbib}
%\bibliographystyle{IEEEtran}
\bibliography{template}  %%% Uncomment this line and comment out the ``thebibliography'' section below to use the external .bib file (using bibtex) .

%%% Uncomment this section and comment out the \bibliography{references} line above to use inline references.
% \begin{thebibliography}{1}

% 	\bibitem{kour2014real}
% 	George Kour and Raid Saabne.
% 	\newblock Real-time segmentation of on-line handwritten arabic script.
% 	\newblock In {\em Frontiers in Handwriting Recognition (ICFHR), 2014 14th
% 			International Conference on}, pages 417--422. IEEE, 2014.

% 	\bibitem{kour2014fast}
% 	George Kour and Raid Saabne.
% 	\newblock Fast classification of handwritten on-line arabic characters.
% 	\newblock In {\em Soft Computing and Pattern Recognition (SoCPaR), 2014 6th
% 			International Conference of}, pages 312--318. IEEE, 2014.

% 	\bibitem{hadash2018estimate}
% 	Guy Hadash, Einat Kermany, Boaz Carmeli, Ofer Lavi, George Kour, and Alon
% 	Jacovi.
% 	\newblock Estimate and replace: A novel approach to integrating deep neural
% 	networks with existing applications.
% 	\newblock {\em arXiv preprint arXiv:1804.09028}, 2018.

% \end{thebibliography}

\end{document}